\documentclass{iopjournal}

\usepackage{graphicx} 
\usepackage{amsmath} 
\usepackage{cleveref} 

\usepackage[normalem]{ulem}
\usepackage{lmodern}

\usepackage{bm}
\usepackage[mathlines]{lineno}
\usepackage{xcolor,soul}
\usepackage{float}

\usepackage{cite} 

\usepackage{booktabs}

\newcommand{\bmvec}[1]{\bm{\mathrm{#1}}}
\newcommand{\deriv}[2]{\frac{\mathrm{\partial}#1}{\mathrm{\partial}#2}+\bmvec{V}\cdot\nabla#1}
\newcommand{\simcode}{M3D-C1}

\begin{document}


\title{Simulations of internal kink modes and sawtooth crashes for SPARC baseline-like scenarios using the \simcode{} code}

\author{W.H. Wang$^1,^*$\orcid{0000-0001-8743-0430}, C. Clauser$^1$\orcid{0000-0002-2597-5061}, C. Liu$^2$\orcid{0000-0002-6747-955X}, N. Ferraro$^3$\orcid{0000-0002-6348-7827}, and R. A. Tinguely$^{1}$\orcid{0000-0002-3711-1834}}

\affil{$^1$Plasma Science and Fusion Center, Massachusetts Institute of Technology, MA 02139, USA}

\affil{$^2$Peking University, Beijing 100871, China}

\affil{$^3$Princeton Plasma Physics Laboratory, NJ 08543, USA}

\affil{$^*$Author to whom any correspondence should be addressed.}

\email{wenhaw42@psfc.mit.edu}


\begin{abstract}
A relaxed baseline case, based on the Primary Reference Discharge (PRD) design point, is used to conduct a first dedicated investigation for the most unstable low-$n$ MHD instabilities in SPARC baseline-like scenarios. The simulations use the high-fidelity 3D extended-MHD code \simcode{}, focusing on a qualitative understanding of potential sawtooth crashes. The linear simulation, by scanning over the resistivity, identifies a dominant internal kink mode at the $q=1$ surface with a toroidal mode number $n=1$. Both the current and the pressure profiles are strongly affecting the kink instability in the baseline case. The linear growth rate is sensitive to the keV-level temperature profile and the on-axis $q_0$ around unity. A simplified 1D eigenvalue solver provides additional support for elucidating the physical mechanisms of the 1/1 mode and qualitative interpretations for the M3D-C1 results. In 3D nonlinear simulations, the marginally unstable case gives a moderate sawtooth crash when $q_0$ is below unity. When both the current and the pressure drives exist (the baseline case), a strong sawtooth crash is obtained, which features a magnetic reconnection event and a hollowed pressure profile. These are consistent with features from both the Kadomtsev and Wesson models. 
The sawtooth oscillations shown in low-$\beta_0$ simulations also provide an initial effort to study sawtooth evolution in SPARC. However, several important physical effects are not yet included, such as two-fluid effects, kinetic alpha particles, and the associated heating sources, limiting quantitative predictions of crash amplitude and timescales. The present work therefore provides a qualitative physics basis for future studies of particle and heat transport during MHD activity in SPARC and future fusion pilot plants.


\end{abstract}




\section{Introduction}




\par
SPARC is a high-field, high-current compact tokamak designed jointly by the MIT Plasma Science and Fusion Center (MIT PSFC) and the Commonwealth Fusion Systems (CFS).\cite{Creely2020} The goal of the device is to pave the road for the fusion pilot plant, ARC, to generate electric power for commercial uses.\cite{Hillesheim2026} One major mission for SPARC is to demonstrate $Q>1$ plasma conditions, where $Q$ is a ratio of the total fusion power to the external power absorbed into the plasma. For a burning plasma condition of the self-sustaining fusion, higher $Q$ values ($Q>5$) are required.\cite{Creely2020} To achieve the predicted $Q\sim 10$ performance for SPARC, the ion cyclotron range-of-frequencies (ICRH) heating and the alpha heating should be accounted for to sustain the plasma in the H-mode with the operational density and temperature.\cite{Scott2020,Creely2020,Rodriguez-Fernandez2020} Therefore, understanding the alpha particle transport in the device is one of the central interests for success. One concern is the transport associated with MHD instabilities that may occur in the SPARC tokamak, since the MHD instability can cause drastic plasma changes, leading to potential losses of alpha particles.
\par
During high fusion power operations of SPARC, a large number of high-energy alpha particles are generated in the core region.\cite{Scott2020,Creely2020} Given the high performance parameters of SPARC, a better understanding of interactions between the alpha particles and various MHD activities can be important to achieve the desired burning plasma scenarios. Such studies can also provide meaningful understandings of many MHD activities, such as sawtooth oscillations and disruption events, which are actively investigated in many tokamak experiments. 
\par
As a first step, a thorough evaluation of the MHD stability is necessary for SPARC operations. An overall discussion of various MHD activities in SPARC can be found in Sweeney et al. 2020.\cite{Sweeney2020} Since the safety factor q-profile soon drops below 1 after the beginning of the flat-top, sawtooth crashes are predicted for SPARC using TRANSP with the Porcelli model. \cite{Porcelli1996, Rodriguez-Fernandez2020} Such sawtooth oscillations may cause strong redistribution of fast ions and thus change the heating in the tokamak. Potential transport of fast ions can also happen by introducing additional modes into the system, leading to particle losses towards the first wall. Particle losses also reduce the heating supplied by alpha particles, which is necessary for maintaining high-Q performance of the SPARC tokamak. Moreover, the onset of the crash may also lead to additional MHD modes in the long term, 
such as tearing modes, or even disruptions.
Though not included in this paper, a possible approach for investigating alpha particle transport is to perform particle-in-cell (PIC) simulations for these kinetic alpha particles with a specified distribution, such as a slowing-down distribution. This allows us to study the interaction between energetic alpha particles and the evolving electromagnetic field in a tokamak. Such studies are planned for future work and will be reported in subsequent publications.
\par
A qualitative sawtooth simulation is carried out by using the TRANSP code with the Porcelli model.\cite{Rodriguez-Fernandez2020} The model calculates the instability onset threshold to decide when to perform a profile crash. The model uses a magnetic reconnection fraction factor to control the radial range of profile flattening around the $q=1$ flux surface, and the parameters were tuned in previous studies that best fit the JET and TFTR experiments.\cite{Bateman2006} Since the extent of the profile flattening and the sawtooth oscillation period heavily depend on the fraction number, the estimation of the crash period may be kept for further investigation. Therefore, attempts to establish reliable sawtooth modeling for SPARC can be beneficial in confirming these events early in the design of high-performance scenarios.
\par
A first challenge is to identify the triggering mechanism of a sawtooth crash in the SPARC tokamak. Several existing theories for the onset of sawtooth crash can be considered for the SPARC simulations. Kadomtsev (1975)\cite{Kadomtsev1975} first proposed a magnetic reconnection model. The peaked current results in a $q$ profile below 1, triggering a 1/1 instability that initiates the sawtooth crash. Additional modifications to this model, such as neoclassical resistivity\cite{Park1990}, two-fluid effects\cite{Wesson1990,Porcelli1991,Zakharov1993,Gunter2015}, etc., have been suggested to explain the fast timescale of the profile collapse in experiments. An alternative Wesson model (1986)\cite{Wesson1986} is used to explain the JET experiments which seem to conflict with the Kadomtsev interpretation. Jardin, et al.\cite{Jardin2020} extend the Wesson model by proposing that a 1/1 mode keeps a flat $q$ just above unity which then excites higher $m=n$ MHD interchange modes to cause a sudden sawtooth crash. To check whether these theories can be applied to a sawtooth crash event in the SPARC, a thorough investigation of 3D nonlinear MHD simulations is needed using SPARC-related equilibria. 
\par
The \simcode{} code\cite{Breslau2009} can be used to carry out the sawtooth study in the SPARC. It is a high-fidelity extended-MHD code which has been used for many sawtooth modeling studies in the past. A multi-timescale sawtooth oscillation is simulated in some early works using the \simcode{}, where periodic complete magnetic reconnection events can be captured when applying a loop voltage at the boundary to maintain the plasma current.\cite{Jardin2012} Later on, the code has also been used to study the non-sawtoothing discharges, due to a ``flux-pumping'' mechanism.\cite{Jardin2015} A follow-up 3D nonlinear study further proves its capability of capturing a saturated quasi-interchange instability responsible for the dynamo generation for the flux pumping.\cite{Krebs2017} More recent works use the \simcode{} to extend the Wesson model for a complete sawtooth oscillation\cite{Jardin2020}, and explain the peaking limitation of the electron temperature in an NSTX discharge.\cite{Jardin2022,Jardin2023} All these simulations prove the \simcode{}'s capability to explore possible sawtooth crash mechanisms in the SPARC tokamak. 
\par
Some preliminary sawtooth crash results have already been reported for the SPARC baseline case in a recent \simcode{} study. Kleiner, et al. (2025)\cite{Kleiner2024} discuss the disruption mitigation using a massive gas injection (MGI) in SPARC. However, this work mainly focuses on the disruption process, while the early stage sawtooth crash activity before the thermal quench is not thoroughly explored by itself. To get a complete picture of such MHD activities that may happen in SPARC, a systematic study of $n=1$ mode using the baseline equilibrium of SPARC is presented in this paper. A better understanding of such an MHD event can provide a guideline for future SPARC operation, and also serve as a solid background to carry out further investigation of interactions between energetic particles and MHD instabilities in SPARC. 
\par
 In this paper, we present the first dedicated M3D-C1 study of low-$n$ MHD instabilities and initial sawtooth-crash dynamics using several baseline-like equilibria derived from the SPARC primary reference discharge (PRD) design point. Because the present study uses a single-fluid extended-MHD model, we focus on the qualitative interpretation of possible sawtooth crash mechanisms in SPARC baseline-like equilibria. Quantitative predictions of sawtooth oscillations in SPARC will require a more complete physics model, including two-fluid effects, kinetic alpha particles and additional heating sources from ICRH and energetic alpha particles. Some key parameters of the baseline equilibrium profiles are discussed in Sec.~\ref{sec:setup}, and we explain the simulation setup using the \simcode{} in this section. In Sec.~\ref{sec:linear}, a dominant $n=1$ mode is observed in the linear \simcode{} simulations using a relaxed SPARC baseline equilibrium. By scanning over the Lundquist number, this mode is identified as an internal kink mode. Both the pressure and the current can have strong effects on this kink mode. The plasma $\beta$ effect is further analyzed by a simplified 1D model. In Sec.~\ref{sec:nonlinear}, we start from a marginally unstable case to simulate possible sawtooth crashes in nonlinear 3D simulations, showing moderate profile flattening with a 20-30\% drop of the on-axis value. When both the pressure and the current drives are included, a stronger flattening can occur for up to 40--50\% of the original central pressure value. The magnetic reconnection and the hollowed pressure profile can be explained by combining the Kadomtsev model and the Wesson model. Sec.~\ref{sec:conclusion} summarizes the simulation results and discusses the future work to properly assess the confinement performance with a better understanding of sawtooth events in SPARC.


%
%



\section{Simulation setup}\label{sec:setup}

\subsection{SPARC primary reference discharge}\label{subsec:eq_prd}

\par
\begin{figure}[t]
   \centering
      \begin{minipage}[t]{0.6\textwidth}\centering
      \includegraphics[width=\textwidth]{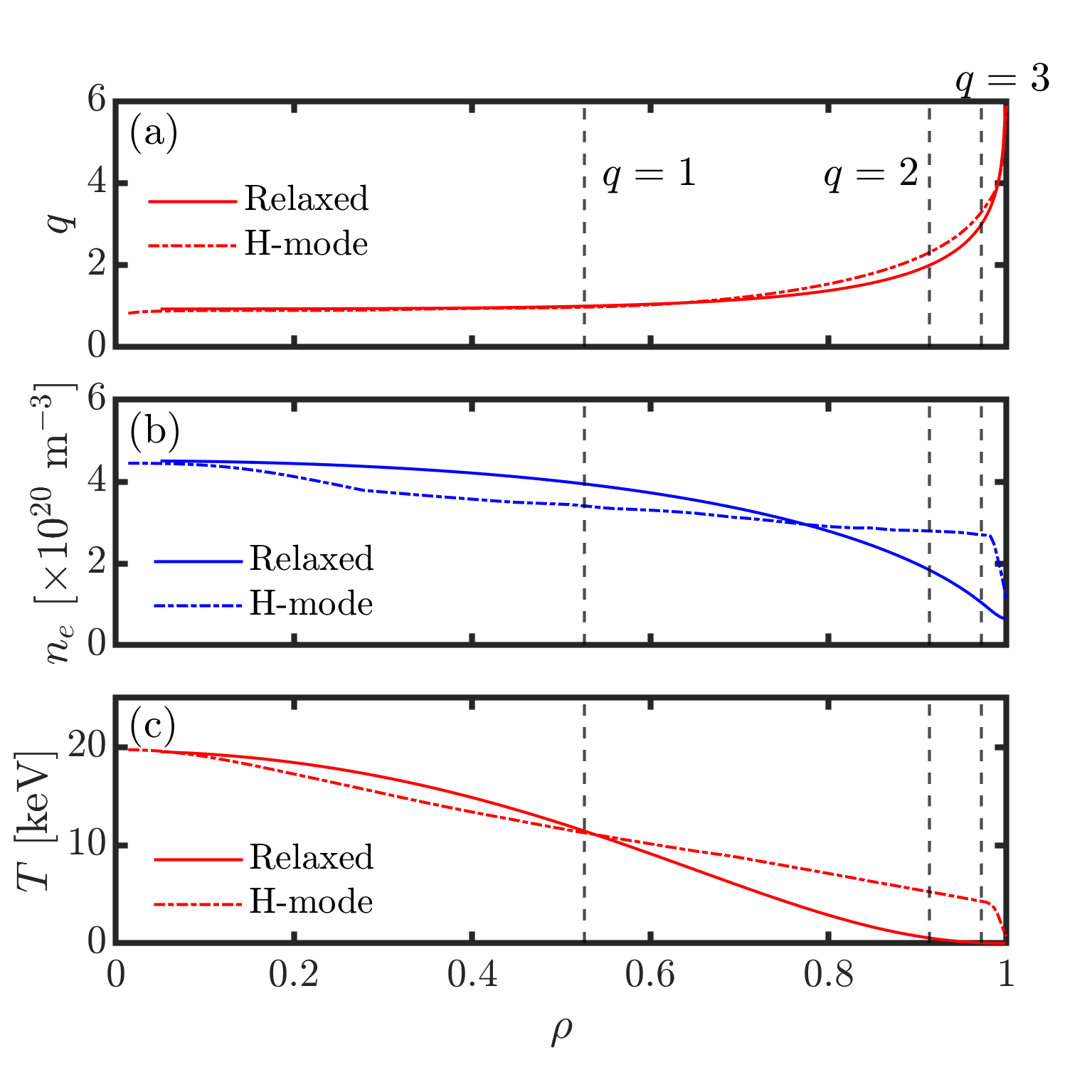}
      \end{minipage}
   \caption{Relaxed equilibrium (labeled ``Relaxed") profiles based on SPARC PRD design point: Panel (a) is the safety factor $q$, where the $\rho_1=0.53$, $\rho_2=0.91$ and $\rho_3=0.97$ are $q=1,2,3$ locations respectively (labeled with vertical dashed lines). Panel (b) is the density profile $n_i=n_e$. Panel (c) is the temperature profile, for SPARC baseline case in this paper $T_i=T_e$. The PRD H-mode profiles (dashed lines labeled ``H-mode") are also provided for reference. All profiles are functions of the radial coordinate, $\rho=\sqrt{(\psi-\psi_0)/\psi_{\mathrm{max}}}$, where the poloidal flux $\psi$ is normalized by its maximum magnitude $\psi_{\mathrm{max}}=\psi_w-\psi_0$ between the LCFS and the magnetic axis.}
   \label{fig_eq_prof}
\end{figure}

The primary reference discharge (PRD) for SPARC is a designed H-mode scenario which is meant to achieve $Q>10$ performance. The profiles of the PRD design point equilibrium are plotted as reference values (labeled ``H-mode") in Fig.~\ref{fig_eq_prof}.\cite{Rodriguez-Fernandez2022b,Rodriguez-Fernandez2022} The baseline equilibrium in our study is a relaxed version of the PRD design point computed by the cfsPOPCON\cite{Body2023}, which will be discussed in detail in this section, and will be referred to as the ``baseline" case in this paper. 
\par
The choice of the relaxed profiles eliminates edge localized instabilities due to steep gradients near the boundary. Because the \simcode{} nonlinear simulation allows free evolution of plasma profiles, the simulation with a pedestal requires a clear understanding of edge physics to maintain the H-mode confinement self-consistently, which is beyond the scope of this paper. Therefore, to avoid unnecessary effects, we use the relaxed PRD profiles shown in Fig.~\ref{fig_eq_prof} as our baseline case.
\par
The modified profiles are qualitatively equivalent to the H-mode scenario within the $q=1$ flux surface. Given our interest in the internal MHD mode in the central core region, the equilibrium should be sufficient to provide insights on the inner core physics of SPARC. This relaxed equilibrium, which removes the pedestal region with steep gradients, has also been used in a previous study for disruption mitigation with massive gas injection.\cite{Kleiner2024} 
\par
The equilibrium has a peaked density on axis, $n_{\mathrm{max}} = 4.5\times 10^{20}$~m$^{-3}$ and the on-axis temperature is $T_{e,0} = T_{i,0} = 20$~keV. The radial profiles of density $n_e$, and temperature, $T_e$ and $T_i$, are shown in Fig.~\ref{fig_eq_prof} as a function of normalized poloidal flux, $\rho = \sqrt{(\psi-\psi_0)/\psi_{\mathrm{max}}}$, with $\psi_{\mathrm{max}}=\psi_w - \psi_0$ being the flux difference between the last closed flux surface (LCFS) and the magnetic axis.   

\par
The on-axis toroidal magnetic field is $B_0 = 12.2$ T, with a plasma current $I_p=8.7$ MA. The safety factor $q$ profile in the baseline case is also provided in Fig.~\ref{fig_eq_prof}, with a $q=1$ location at $\rho_1=0.53$. Since there exists a considerable volume of plasma that has $q<1$ for $\rho<\rho_1$, we can expect low-$n$ MHD instabilities to occur in the core region. Higher rational surfaces, such as $q=2$ and $q=3$, are also presented in the baseline equilibrium near the edge (beyond $\rho>0.8$). This provides a possibility for the mode at $q=1$ surface to be coupled to higher rational $q$ surfaces when forming a global mode structure, which will be shown in our discussion in the following sections. The low-$n$ MHD instabilities may lead to drastic profile changes, hence affecting the performance of SPARC operation. Thus, understanding the linear and nonlinear properties of such MHD activities in SPARC can be an essential topic.

\subsection{\simcode{} code and setup for SPARC simulations}\label{subsec:m3dc1}

\par
\simcode{} is a nonlinear 3D extended-MHD code that uses high-order finite elements and implicit time-stepping to advance the equations \cite{Jardin2012, Ferraro2018}. These features allow the use of sufficiently large time steps while maintaining accuracy for long simulations covering different timescales. The code can perform both linear and nonlinear simulations for 3D geometry. The unstructured mesh grids and various physics models also provide flexibility to model arbitrary magnetic geometries with multiple wall regions. This allows us to conduct realistic simulations for SPARC.  
\par
The extended MHD model used in this paper can be summarized as:\cite{Ferraro2018}
\begin{equation}
\label{eq_den}
\frac{\partial n_i}{\partial t} + \nabla \cdot (n_i \bmvec{V}) = \nabla \cdot (D \nabla n_i),
\end{equation}

\begin{equation}
\label{eq_mhd}
n_i M\left[\deriv{\bmvec{V}}{t}\right] = - \nabla P + \bmvec{J}\times\bmvec{B} - \nabla \cdot \Pi,
\end{equation}


\begin{equation}
\label{eq_b_dot}
\frac{\partial \bmvec{B}}{\partial t} = \nabla\times \left(\bmvec{V}\times\bmvec{B} - \eta \bmvec{J} \right), \quad 
\mu_0 \bmvec{J} = \nabla\times\bmvec{B}
\end{equation}

\begin{equation}
\label{eq_te}
n_e \left[ \deriv{T_e}{t}+(\Gamma-1)T_e\nabla \cdot \bmvec{V} \right]= 
(\Gamma-1) \left[\eta J^2 - \nabla\cdot\bmvec{q}_e + Q_e - \Pi_e : \nabla\bmvec{V} \right]
\end{equation}

\begin{equation}
\label{eq_ti}
n_i \left[ \deriv{T_i}{t}+(\Gamma-1)T_i\nabla \cdot \bmvec{V} \right]= 
(\Gamma-1) \left[ - \nabla\cdot\bmvec{q}_i + Q_i - \Pi_i : \nabla\bmvec{V} \right]
\end{equation}
where Ohm's law with a resistive current, $\bmvec{E} = \eta \bmvec{J} - \bmvec{V}\times\bmvec{B}$, is employed for Eq.~\ref{eq_b_dot}. The total pressure $P = P_i+P_e = n_i (T_i + Z_i T_e)$, provided quasi-neutrality, $n_i Z_i= n_e$. The stress tensor $\Pi = \Pi_i+\Pi_e$ refers to the fluid viscous effects, which gives the fluid diffusion, $\nabla\cdot\Pi = \mu\nabla^2\bmvec{V}$.
\par
The \simcode{} code expresses the magnetic field as $\bmvec{B} = \nabla \psi_p \times \nabla\Phi-\nabla_\perp f^\prime + I\nabla \Phi$, where $\psi_p$ is the poloidal flux function, and $I = R B_t$ is the current function. The \simcode{} uses a cylindrical coordinate system, $(R,\Phi,Z)$, with $\Phi$ being the angular coordinate. The fluid velocity is expressed as $\bmvec{V} = R^2\nabla U\times\nabla\Phi+\omega R^2 \nabla\Phi +R^{-2}\nabla_\perp\chi$.  The eight unknown variables are $n_i$, $T_i$, $T_e$, $U$, $\omega$, $\chi$, $\psi$ and $f$, which will be solved by the equation set, Eq.~\ref{eq_den} - Eq.~\ref{eq_ti}. Depending on the purpose of our simulations, we can simulate a reduced model by fixing some of the field quantities.
\par
For SPARC simulations in this paper, some assumptions are used to simplify the model and help interpret the simulation data. First, no external heating is considered, so the heating source $Q_e$ in \simcode{} is simplified to the energy transfer between species, $Q_e = -Q_i = 3m_en_e(T_i-T_e)\frac{\nu_{ei}}{m_i}$. Next, the heat flux term can be expressed as $\bmvec{q}_j = -\kappa_T\nabla_\perp T_j-\kappa_{T,\parallel}\bmvec{b}\nabla_\parallel T_j$, where the parallel gradient in the magnetic field direction is $\nabla_\parallel = \frac{\bmvec{B}}{B}\cdot\nabla = \bmvec{b}\cdot\nabla$. For constant heat conductivities, we can obtain a perpendicular diffusion term, $\kappa_T\nabla^2 T_j$, which will be used to relax the temperature profiles in some of SPARC simulations in the following sections. The Ohmic heating $\eta J^2$ is always included in the nonlinear simulation which causes the current peaking in the long time simulation.
\par
Moreover, the default resistivity model in our \simcode{} simulations is the Spitzer resistivity,
\begin{equation}
\label{eq_eta}
\eta_s = \frac{m_e\nu_{ei}}{n_e e^2} 
\text{,\quad with \quad} 
\nu_{ei}=\frac{4\sqrt{2\pi}e^4Z_i^2 n_i \ln{\Lambda}}{3\sqrt{m_e}T_e^{3/2}}
\end{equation}
where $\nu_{ei}$ is the collisional frequency between electrons and ions. This typically leads to a relatively large resistivity near the SPARC boundary, but in the core region the resistivity is a few orders of magnitude smaller, since $\eta_s \propto T_e^{-3/2}$. This makes the SPARC baseline case approach the ideal MHD limit as we will see in Sec.~\ref{sec:linear}.
\par
In simulations, transport and dissipation processes are controlled by prescribed models of several transport coefficients. The particle diffusivity $D$ and viscosity $\mu$ determine the diffusion of particle density and velocity respectively. The heat conductivity, $\kappa_T$, affects the temperature relaxation. The evolution of magnetic flux and current is governed by the resistivity $\eta$. In this work, $D$, $\mu$ and $\kappa_{T}$ are prescribed constants in each simulation. $\eta$ is modeled either as a constant or using Spitzer resistivity depending on simulation purposes. For a quantitative analysis of sawtooth oscillation evolution, particularly its characteristic timescales, physically appropriate models and values for these coefficients can be crucial. In the present work, however, we focus on elucidating the physical mechanisms underlying sawtooth crash in SPARC baseline-like scenarios. A more systematic investigation of transport models and their impacts is left for future work.
\par
Typical values of transport coefficients are $D=3.08$ m$^2$/s, $\mu = 1.03\times10^{-4}$ kg/(m$\cdot$s), and $\kappa_T = 1.54\times 10^{20}$ /(m$\cdot$s) with $\kappa_{T,\parallel}\sim 10^6\kappa_T$. Spitzer resistivity $\eta_s$ is used for the resistive effects except for the parameter scan of the Lundquist number. As a reference value, we have $\eta_s=2.56\times 10^{-9}$~Ohm$\cdot$m at the $q=1$ flux surface, $\eta_s=3.39\times 10^{-7}$~Ohm$\cdot$m at $q=2$ and $\eta_s=3.28\times 10^{-5}$~Ohm$\cdot$m at $q=3$. Clearly, there is a rapid increase of the resistivity due to the large drop of the pressure near the boundary. As we will see in the next section, the dominant mode appears at $q=1$ surface with a relatively low $\eta_s$, with an extended structure at higher rational $q$ flux surfaces.

\begin{figure}[t]
   \centering
   \begin{tabular}{cc}
      \begin{minipage}[t]{0.585\textwidth}\centering
      \includegraphics[width=\textwidth]{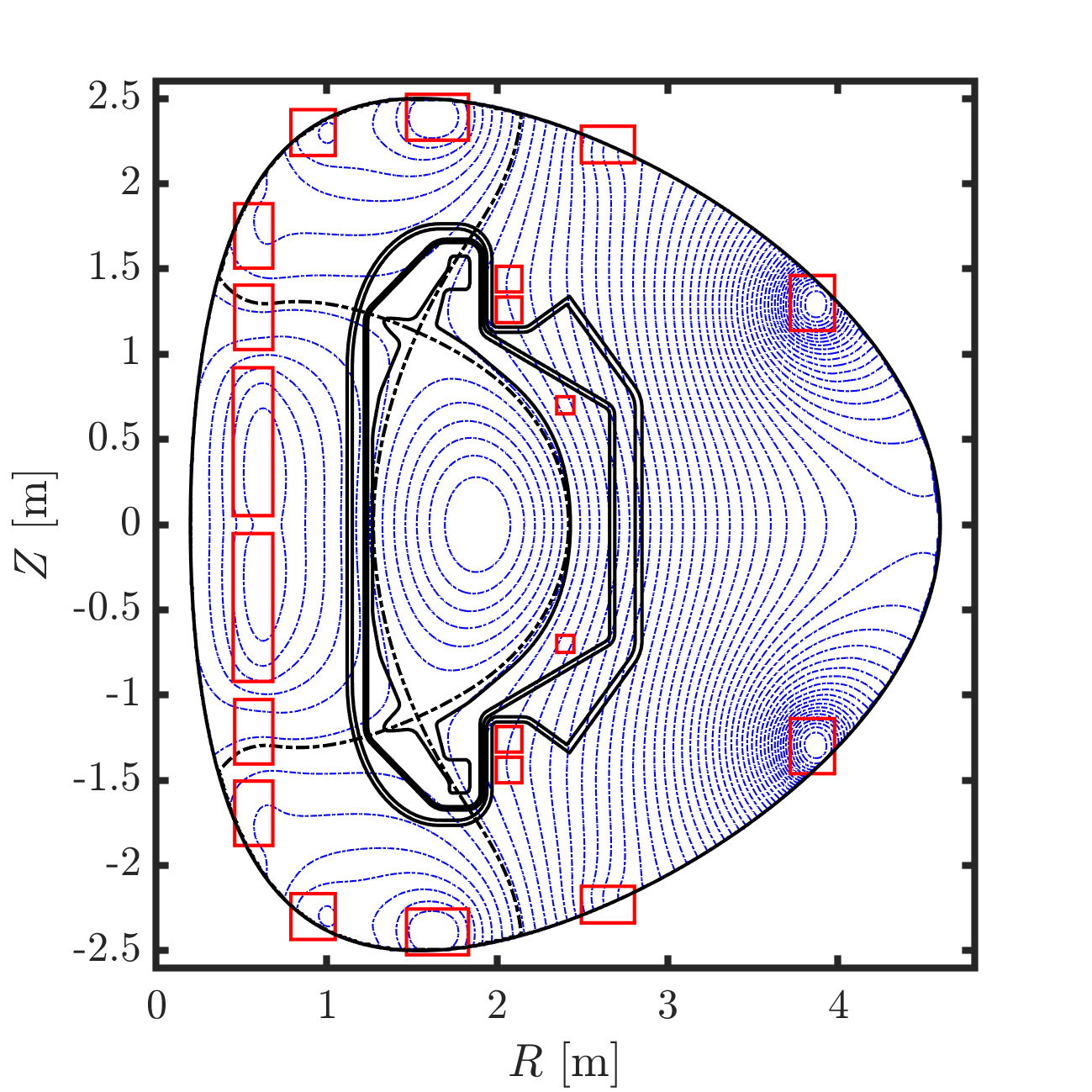}
      (a) The whole device geometry
      \end{minipage}
      
      \begin{minipage}[t]{0.375\textwidth}\centering
      \includegraphics[width=\textwidth]{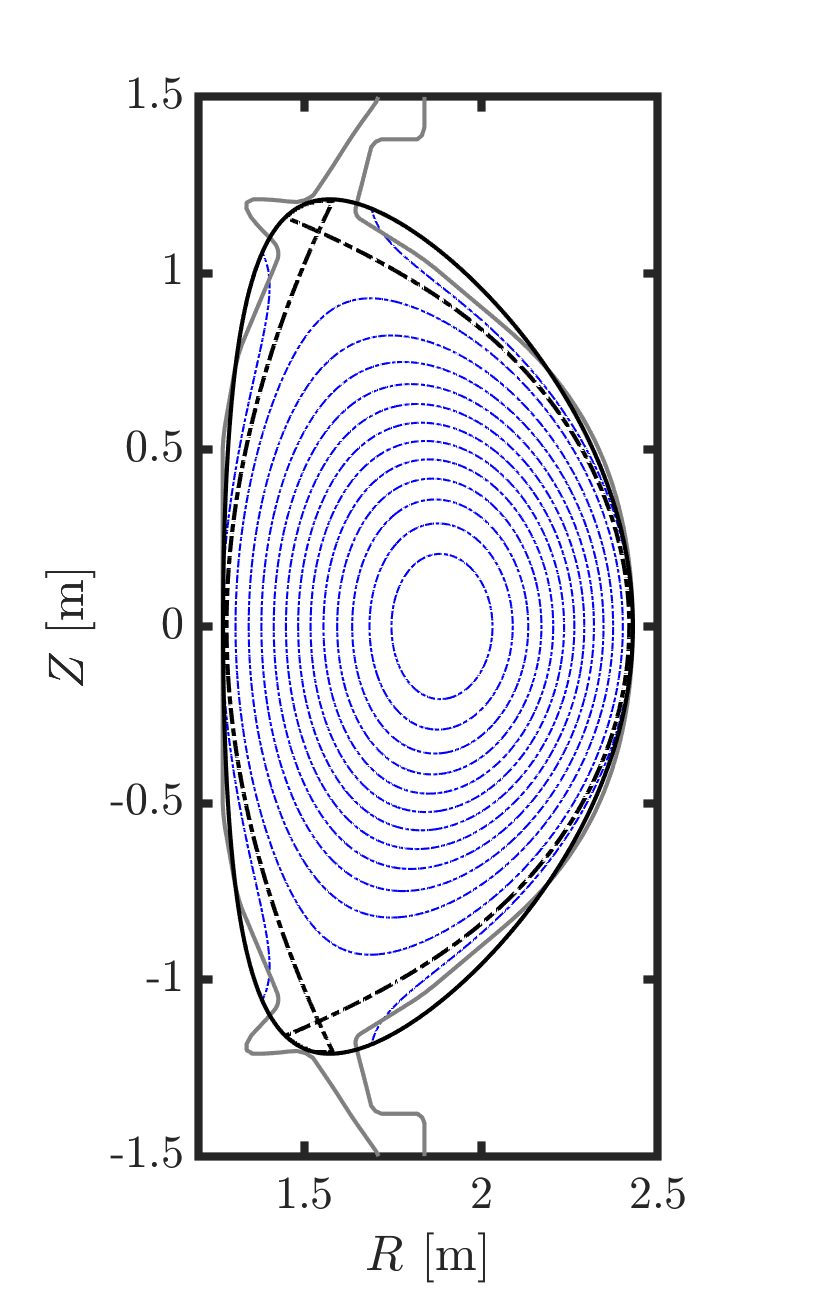}  
      (b) The simple geometry
      \end{minipage}
   \end{tabular}
   \caption{Mesh geometries used in \simcode{} simulations: (a). high resolution mesh region including the first wall and the conducting vessel walls (solid lines). The coil locations are marked in red rectangles and the poloidal flux geometry is shown by the dashed blue contours. The LCFS is shown by the black dashed line. (b). simplified mesh region with plasma region only. The actual simulation boundary (black solid line) is different from the first wall (gray solid line) but contains a comparable plasma volume. The blue contours are the poloidal flux surfaces and the LCFS is the black dashed line. }
   \label{fig_mesh_grid}
\end{figure}
\par
Two sets of mesh grids are used in this study, shown in Fig.~\ref{fig_mesh_grid}. The cylindrical coordinates, $(R,\Phi,Z)$, are used as a global coordinate system throughout the paper. The first mesh (Fig.~\ref{fig_mesh_grid}a) includes all the vessel walls and the coils to generate a self-consistent equilibrium. The mesh grid contains around 56k elements in total. The outermost simulation boundary imposes a no-flow condition with a perfectly conducting boundary condition. The vessel wall geometries inside the simulation domain are modeled as resistive walls with low resistivity. A representative value of stainless steel resistivity is used, i.e., $7.46\times 10^{-7}$~Ohm$\cdot$m. The space between the conducting walls is assumed to be vacuum. This mesh will be used for the benchmark case, where we include all the physics in our simulations to make a high-fidelity self-consistent prediction for MHD activities. 
\par
The second mesh (Fig.~\ref{fig_mesh_grid}b) is constructed based on the first mesh geometry. This simple mesh focuses only on the plasma region, with a total of 15k mesh elements. This mesh is used for theoretical analysis and several systematic parameter studies to isolate various physical effects. Since we removed complicated wall geometries in this mesh, the instabilities are usually less noisy when scanning over the parameter space. However, we should note that this simplified mesh will impose a no-flow condition and a conducting wall boundary close to the plasma edge.
\par
Linear \simcode{} simulations assumed a single toroidal mode number. Toroidal mode coupling only appears in the nonlinear \simcode{} simulations, where multiple toroidal modes are included for a self-consistent nonlinear time evolution. In \simcode{}, a reduced quintic triangular element in ($R,Z$) plane and a Hermite cubic element in toroidal $\Phi$ direction are used in 3D simulations.\cite{Jardin2012} The PRD benchmark case is simulated using the highest resolution with $n_t=8$ toroidal planes (which resolves toroidal modes up to $n=7$). For subsequent parameter studies to understand the baseline results, we only use $n_t=4$ toroidal planes (up to $n=3$) to track the minimum required physics in nonlinear evolution. Effects of higher toroidal components are discussed for the baseline case in Sec.~\ref{sec:nonlinear}.



\section{Linear simulation of \texorpdfstring{$n=1$}{n=1} mode in SPARC}\label{sec:linear}




\begin{figure}[t]
   \centering
   \begin{tabular}{cc}
      \begin{minipage}[t]{0.425\textwidth}\centering
      \includegraphics[width=\textwidth]{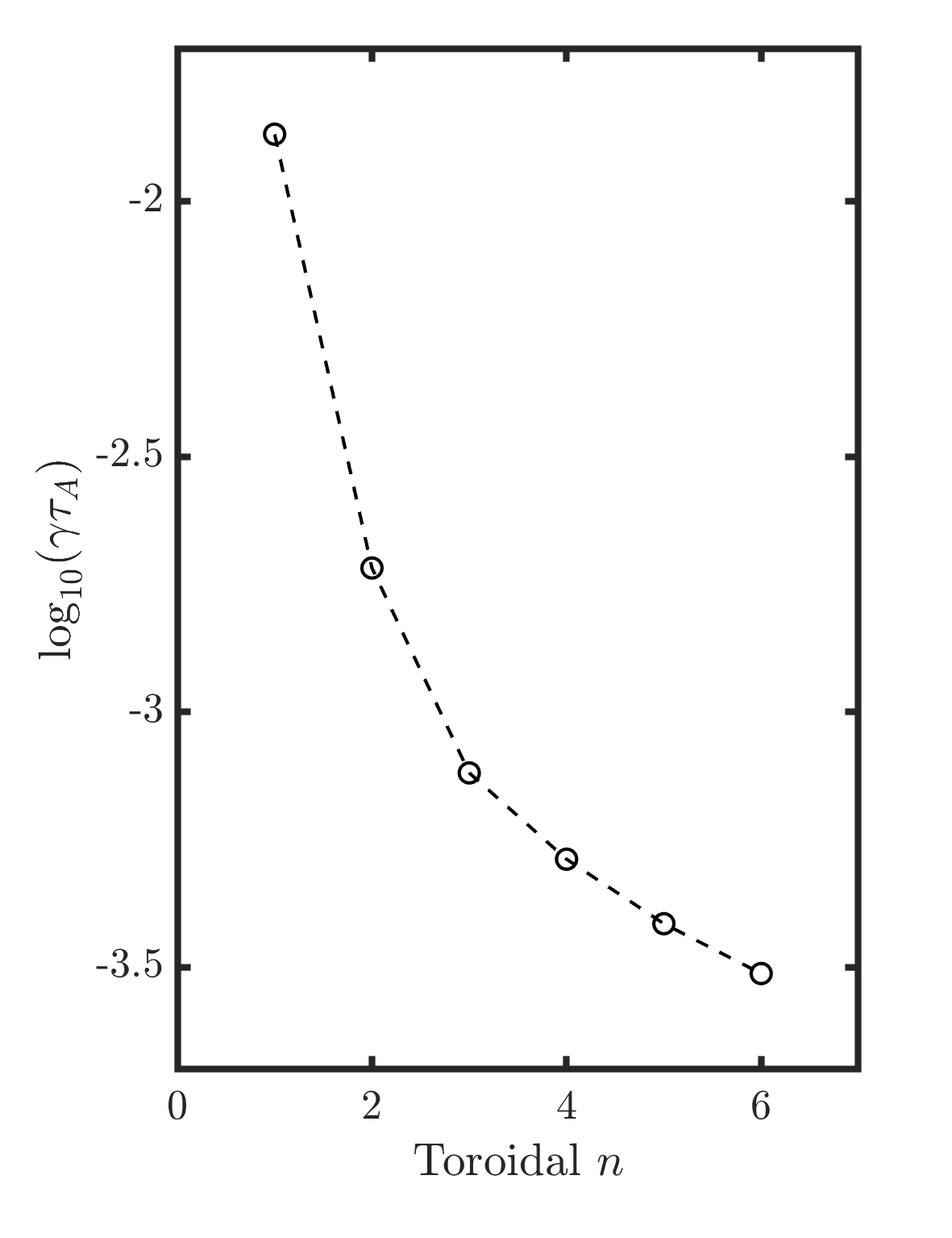}
      (a) Linear growth rate
      \end{minipage}
      
      \begin{minipage}[t]{0.48\textwidth}\centering
      \includegraphics[width=\textwidth]{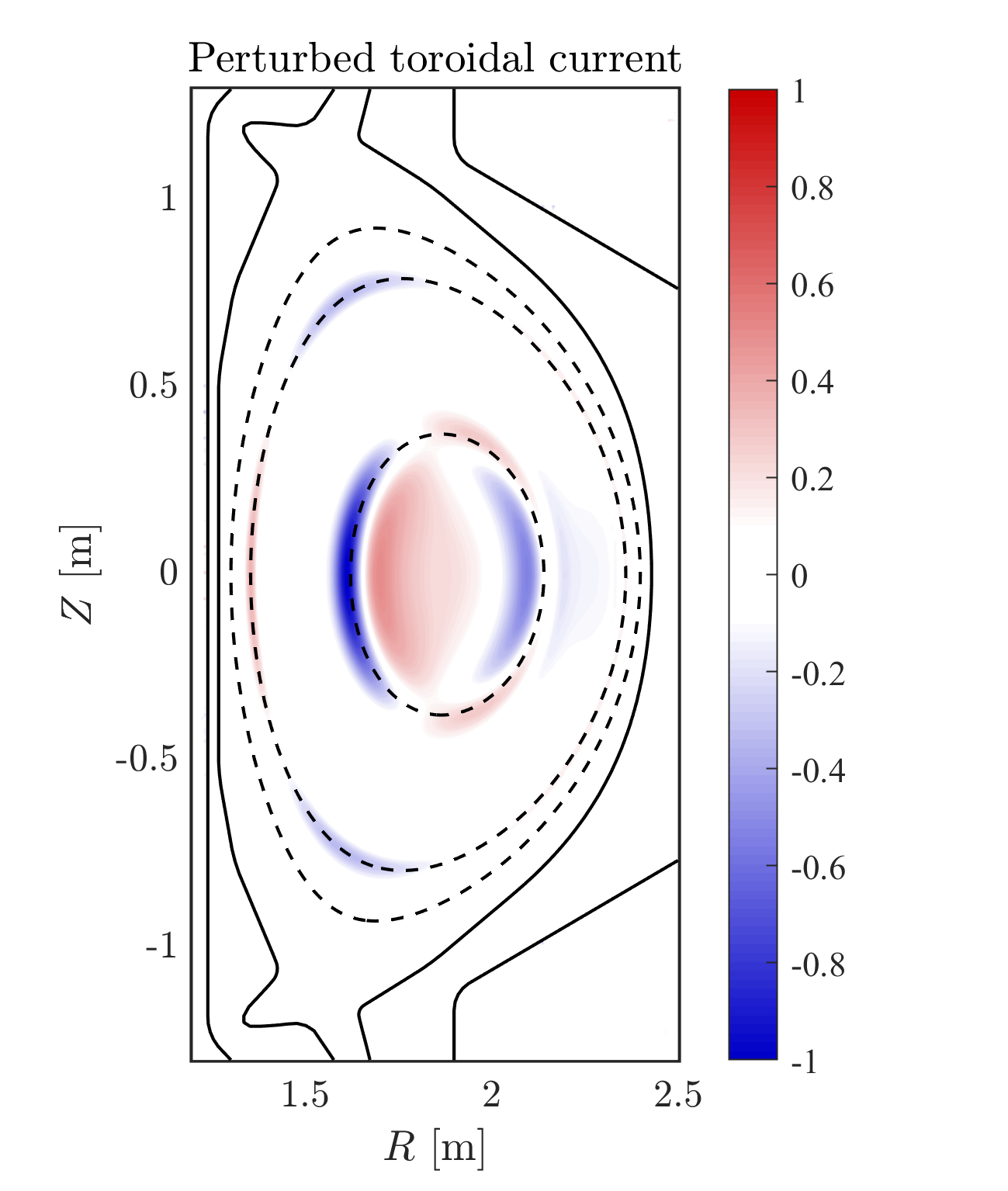}
      (b) Poloidal mode structure
      \end{minipage}
   \end{tabular}
   \caption{(a) Linear growth rate $\gamma$ for a single toroidal mode $n$. The simple geometry is used in the toroidal mode scan.} The dashed line is a fitting line to show the decreasing trend of $\gamma$. The normalization is $\tau_A = 2.10\times 10^{-7}$ s. (b) Poloidal mode structure of the perturbed toroidal current $\delta j_\Phi$ for $n=1$ mode in the baseline equilibrium. The whole device geometry is used to get an accurate mode structure. The dashed lines indicate the flux surfaces at $q=1,2,3$. The linear mode is normalized to its maximum amplitude. The solid lines are schematic contours of the first wall and the vacuum vessel in the whole-device \simcode{} simulations.
   \label{fig_ntor_scan}
\end{figure}
\par
In this section, we will begin with the linear simulation to find the most unstable low-$n$ mode and its characteristics for the SPARC baseline case. To have a clear understanding, we carried out several parametric studies, such as scanning over the resistivity, the plasma pressure and safety factor $q$ profile. These linear results provide the basis for subsequent nonlinear simulations of sawtooth crashes in the next section.
\par
Toroidal mode numbers, $n=1\sim 6$, are simulated separately to find the most unstable mode and linear growth rates for each toroidal $n$ are shown in Fig.~\ref{fig_ntor_scan}a. The linear growth rate is normalized by the Alfv\'en time $\tau_A = R_0/V_A = 2.10\times 10^{-7}$ s, with $V_A = B_0/\sqrt{\mu_0 m_i n_i}$. Here, the magnetic axis location is $R_0 = 1.85$ m, and the on-axis magnetic field $B_0 = 12.2$~T. A representative on-axis density is $n_0 = 4.5\times 10^{20}$ m$^{-3}$. The mesh grid in these simulations only includes the plasma region to avoid numerical issues at higher toroidal mode numbers, since the growth rate becomes 1--2 orders of magnitude smaller as $n$ increases. 
\par
The ion species we currently use in our simulations is assumed to be Deuterium for a single ion species simulation, i.e., $m_i = 2 m_p$, where $m_p$ is the proton mass. For a simplified MHD formulation with $T_i=T_e$, the ion mass effect in Eq.~\ref{eq_den}--\ref{eq_ti} generally becomes a scaling factor in the time unit, and the dimensionless form of equations will not be affected. For a 50:50 mixture of Deuterium and Tritium as required to achieve the PRD design point, the time scaling factor is 1.12 in our simulations, with a coarse estimate of $m_i = 2.5 m_p$. However, a complete study of isotope effects should be considered in future work, including both Deuterium and Tritium as separate ion species in the simulation. This may modify collisions and resistivity calculations due to multiple ion species. In addition, the resulting energetic alpha particle population will contribute to the total pressure and influence the mode growth and nonlinear evolution of bulk plasma.
\par
Fig.~\ref{fig_ntor_scan}a indicates that an $n=1$ mode has the largest linear growth rate of $\gamma\tau_A = 1.36\times10^{-2}$. The toroidal mode scan is completed using the simple mesh for plasma region only (Fig.~\ref{fig_mesh_grid}b). When using the whole device mesh, the linear growth rate gets slightly larger, $\gamma\tau_A = 1.67\times10^{-2}$. This is because the simulation domain boundary in the whole device mesh is far from the plasma region, and the resistive vessel walls set the plasma boundaries. For the simple geometry, the perfectly conducting and no-flow conditions are imposed closer to the plasma edge and tend to reduce the growth rate a little. Despite these differences, the qualitative characteristics of the mode remain unchanged, given that the mode is dominantly within the plasma core region. The mode structure and the poloidal spectrum are comparable with different geometry, except that the perturbation close to the edge is slightly reduced with the simple mesh due to the stabilizing effect of the conducting boundary condition.
\par
A convergence test is also performed for the mesh resolution: for the simple geometry, mesh grids of $\sim 65$k elements are compared to the $\sim 15$k mesh grids; for the whole device geometry, meshes with $\sim 36$k, $\sim 50$k and $\sim 127$k elements are tested. Both geometries show less than 0.2\% variations in the calculated growth rates. 
\par
For $n=1$, the corresponding linear mode structure in the poloidal plane is visualized in terms of the perturbed toroidal current, $\mathrm{d}j_\Phi$, shown in Fig.~\ref{fig_ntor_scan}b. A dominant $m=1$ component can be seen within $q=1$ surface, and higher $m$ harmonics were also observed between higher rational surfaces: for example, $m=2$ component between $q=1,2$ flux surfaces, and $m=3$ components between $q=2,3$ flux surfaces. These results can also be explained by the nature of the internal kink mode structure (see Sec.~\ref{subsec:beta_model_1d}). In the following discussion, we focus on the most unstable $n=1$ mode and study the linear properties of the observed instability.

\subsection{Internal kink mode in SPARC baseline case}\label{subsec:eta_scan}

\par
\begin{figure}[t]
   \centering
      \begin{minipage}[t]{0.6\textwidth}\centering
      \includegraphics[width=\textwidth]{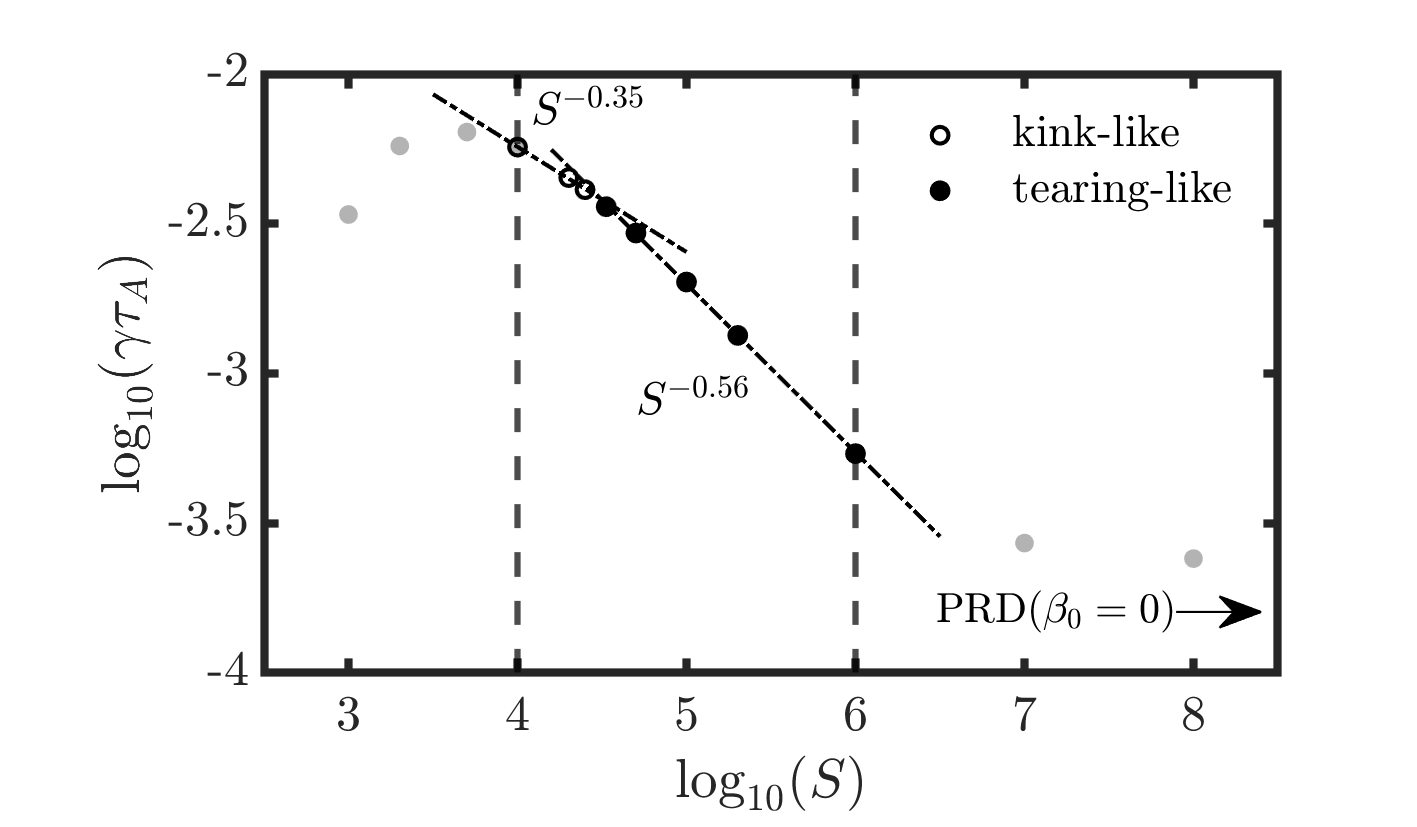}
      \end{minipage}
   \caption{Linear growth rate $\gamma$ with respect to the Lundquist number $S$, by scanning over a constant resistivity $\eta$. The time unit is $\tau_A = 2.10\times 10^{-7}$ s. The relaxed baseline (PRD) case is labeled far to the right of the figure (ideal MHD limit) considering its typical resistivity level in the core region. The vertical lines mark the region where two scaling trends can be compared to the kink mode theory with plasma $\beta_0=0$. The grey data points are beyond the linear fit for scaling, which approach the extremely resistive case (to left) and the ideal MHD case (to right). The dashed fitting lines show qualitative scaling trends. The simple mesh is used to reduce the complicated geometry for a better comparison with the theory.}
   \label{fig_eta_scan}
\end{figure}

To identify the 1/1 mode in our simulation, we scan over the resistivity $\eta$ to study resistive effects on the mode. To make it easier to interpret results, we assume a constant resistivity value for the whole plasma region. By changing the constant $\eta$ value, Fig.~\ref{fig_eta_scan} shows the growth rate scaling with respect to the Lundquist number $S=\frac{V_A R_0}{\eta_{\mathrm{mag}}}$, where the magnetic resistivity is $\eta_{\mathrm{mag}} = \eta/\mu_0$. The mesh grid of a single plasma region is used, because for extremely high constant resistivity, modes at the $q=2$ surface are artificially destabilized when using the whole device mesh. The simple geometry reduces the growth of the $2/1$ mode, because the highly conducting simulation boundary has a stabilizing effect on perturbations close to the edge, e.g., the $q=2$ location. Since such large resistivities are not expected under realistic SPARC parameters, we use the simplified plasma-only mesh for these scans to avoid this artificial complication.
\par 
For the resistivity scan, pressure evolution is artificially suppressed by disabling Eq.~\ref{eq_te} and \ref{eq_ti} in \simcode{}, allowing a more direct theoretical comparison isolating the resistive effects.
The resulting scaling trend is quite similar to the discussion in Hastie, et al. (1987)\cite{Hastie1987} that a kink-like mode structure is obtained at low $S$, whereas a tearing-like mode structure is obtained at high $S$, for $m=1$, $n=1$ mode with the plasma $\beta = 0$. 
\par
In Fig.~\ref{fig_eta_scan}, the linear growth rates within $S\in [3.35\times 10^4,1\times 10^{6}]$ give a linear fit of
\begin{align}
\log_{10}(\gamma\tau_A) = -0.5617\log_{10}(S)+0.1067
\end{align}
This is close to the tearing-like scaling for large $S$, expecting $\gamma \propto S^{-3/5}$. For data points $S\in [1\times 10^4,2.51\times 10^{4}]$, the linear fit gives 
\begin{align}
\log_{10}(\gamma\tau_A) = -0.3512\log_{10}(S)-0.8384
\end{align}
which can be compared to the resistive kink mode scaling, $\gamma \propto S^{-1/3}$. The Lundquist number range of these scaling laws is qualitatively consistent with Hastie's discussion, and the remaining differences may be attributed to geometry changes, because different $q$ profiles, aspect ratios, and a circular cross section were used in Hastie's calculations.\cite{Hastie1987} 
\par
The gray dots in the figure are extreme cases that begin to deviate from the fitted trending lines as shown in Fig.~\ref{fig_eta_scan}. For $S<10^4$, we have very high resistivities of $\eta>1.94\times 10^{-4}$~Ohm$\cdot$m, which is far above the typical Spitzer resistivity for SPARC parameters; for $S>10^6$, we have low resistivities of $\eta<1.94\times 10^{-6}$~Ohm$\cdot$m, which approaches the ideal MHD limit ($\eta \to 0$). Given the PRD designed parameters, most of the core region has a resistivity below $10^{-7}$~Ohm$\cdot$m, which suggests that resistivity effects can be small on linear growth rates. This will be further examined in the plasma $\beta$ scan below. To recover the baseline case, we should consider not only a large Lundquist number of $S\sim 10^9$, but also the pressure effects discussed in the following section.

\par

To explore the stability boundary, the on-axis safety factor, $q_0$, is also scanned over a series of $q$ profiles. A method well-tested in previous \simcode{} simulations to modify $q$ profile is the so-called Bateman scaling\cite{Bateman1977,Jardin2022,Jardin2023}, which keeps the poloidal flux, $\psi$, unchanged by keeping the toroidal current density, $j_\Phi = R p^\prime + R^{-1}II^\prime$, constant. The free function, $I(\psi)=RB_\Phi(\psi)$ can be written as $I(\psi) = \left[I(\psi)-I_{edge}\right] + I_{edge}$, with its edge value, $I_{edge}$, at the separatrix. By scaling the external $I_{edge}$ with a fixed factor, $F_s$, we are able to modify the toroidal magnetic field at the LCFS and hence alter the $q$ profile without changing the poloidal field geometry. 

\begin{figure}[t]
   \centering
      \begin{minipage}[t]{0.6\textwidth}\centering
      \includegraphics[width=\textwidth]{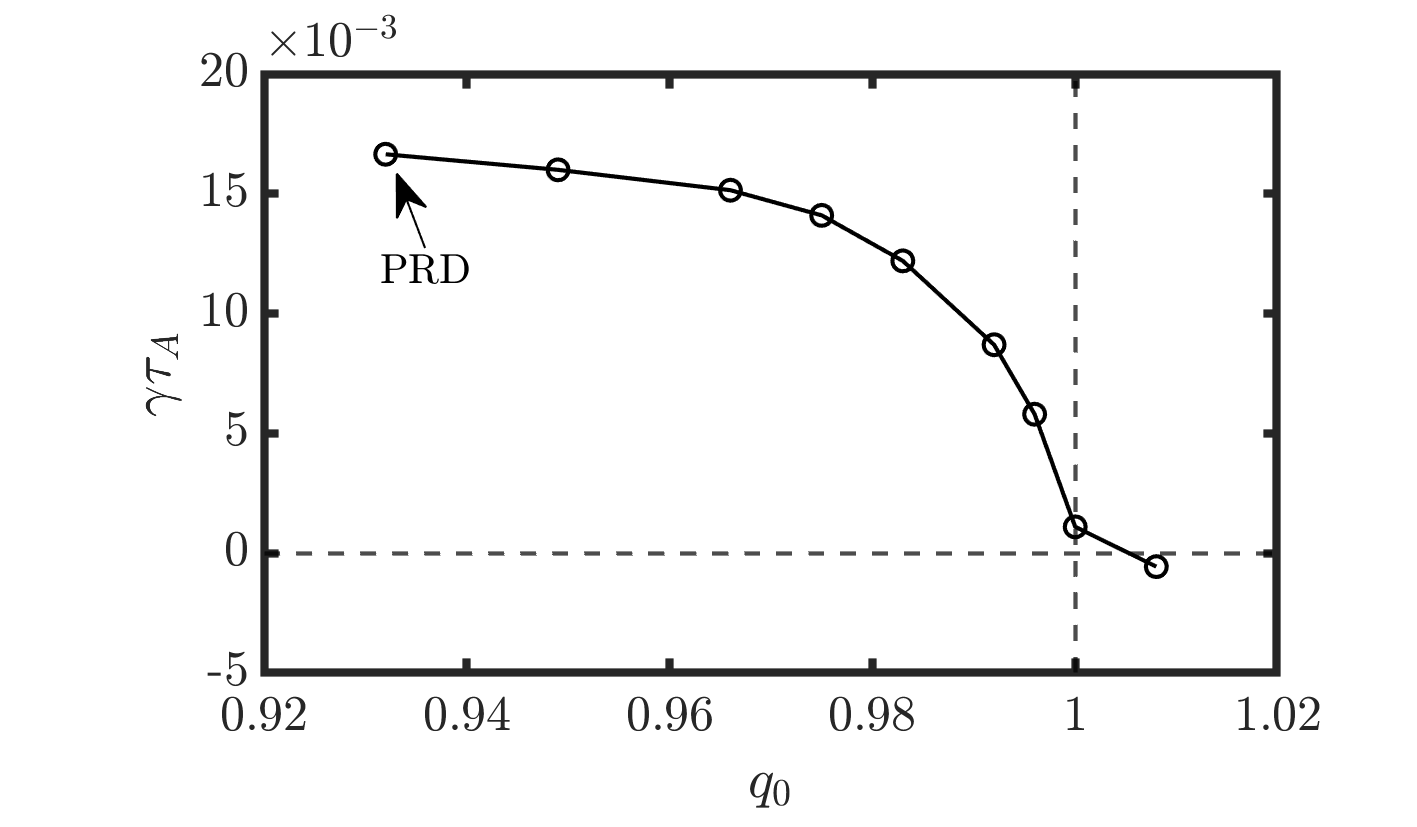}
      \end{minipage}
   \caption{Linear growth rate $\gamma$ with respect to the on-axis safety factor $q_0$, with fixed plasma $\beta_0=4.75$\%. The time unit is $\tau_A = 2.10\times 10^{-7}$ s. The baseline case is labeled as ``PRD". The vertical dashed line marks the $q_0=1$. For the $q_0=1.008$ case, the plasma shows marginal stability. The whole device mesh is used in simulations.}
   \label{fig_q0_scan}
\end{figure}
\par
 During the scan, the on-axis $I_0$ value varies in the range $23.5\sim 25.75$~T$\cdot$m with $F_s=1.0\sim 1.1$. The associated on-axis magnetic field $B_0$ varies from 12.5~T to 13.6~T, causing a less than 10\% change in the Alfv\'en time. As shown in Fig.~\ref{fig_q0_scan}, the linear growth rate can quickly drop by an order of magnitude when the on-axis $q_0$ changes from 0.93 to 1.01. The drastic change happens in a very narrow range around $q_0=$ 0.99 -- 1.01. The onset of the kink mode seems to be earlier than the previous TRANSP simulations\cite{Rodriguez-Fernandez2020}, when $q_0$ is closer to unity. This may be due to the lack of various stabilizing effects that may appear when proper heating sources and kinetic energetic particles, such as alpha particles, are included. In the following nonlinear simulations, it is also shown that the 1/1 mode also changes from a current-driven instability to a pressure-driven instability when $q_0$ approaches 1. The sensitive behavior makes it challenging when predicting the sawtooth crash event for the SPARC design point.


\subsection{\texorpdfstring{Plasma $\beta$}{plasma beta} scan for SPARC baseline case}\label{subsec:beta_scan}

\par
Given the high temperature profiles for SPARC baseline case, it is interesting to study the $\beta$ effects on the kink mode. We define an on-axis plasma $\beta_0 = \frac{2\mu_0 P_{0}}{B_0^2}$, with $P_{0} = n_{0} (T_{i,0}+T_{e,0})$. For the on-axis temperature $T_{i,0}=T_{e,0} = 20$~keV and the density $n_{e,0} = 4.5\times 10^{20}$~m$^{-3}$, we obtain $\beta_0 = 4.75$\%. 
\par 
To scan over plasma $\beta_0$, we artificially relax the equilibrium temperature profile. We set an artificially large thermal heat conductivity, $\bar{\kappa}_T=\frac{\kappa_T}{n_0} \sim \frac{1}{\nabla^2T_e}\frac{dT_e}{dt} = 1.54\times 10^3$ m$^2$/s with a nominal $n_0$ chosen as $1\times 10^{20}$~m$^{-3}$, so that the temperature profiles $T_i$ and $T_e$ are quickly relaxed with the peak value $T_{e,0}$ dropping from $20$~keV ($\beta_0 = 4.75$\%) to around $240$~eV ($\beta_0 = 0.052$\%). The ion temperature $T_{i,0}$ drops in a comparable way from $20$~keV to $140$~eV, since $T_i$ and $T_e$ are evolved separately in the simulation. 
\par
\begin{figure}[t]
   \centering
   \begin{tabular}{cc}
      \begin{minipage}[t]{0.45\textwidth}\centering
      \includegraphics[width=\textwidth]{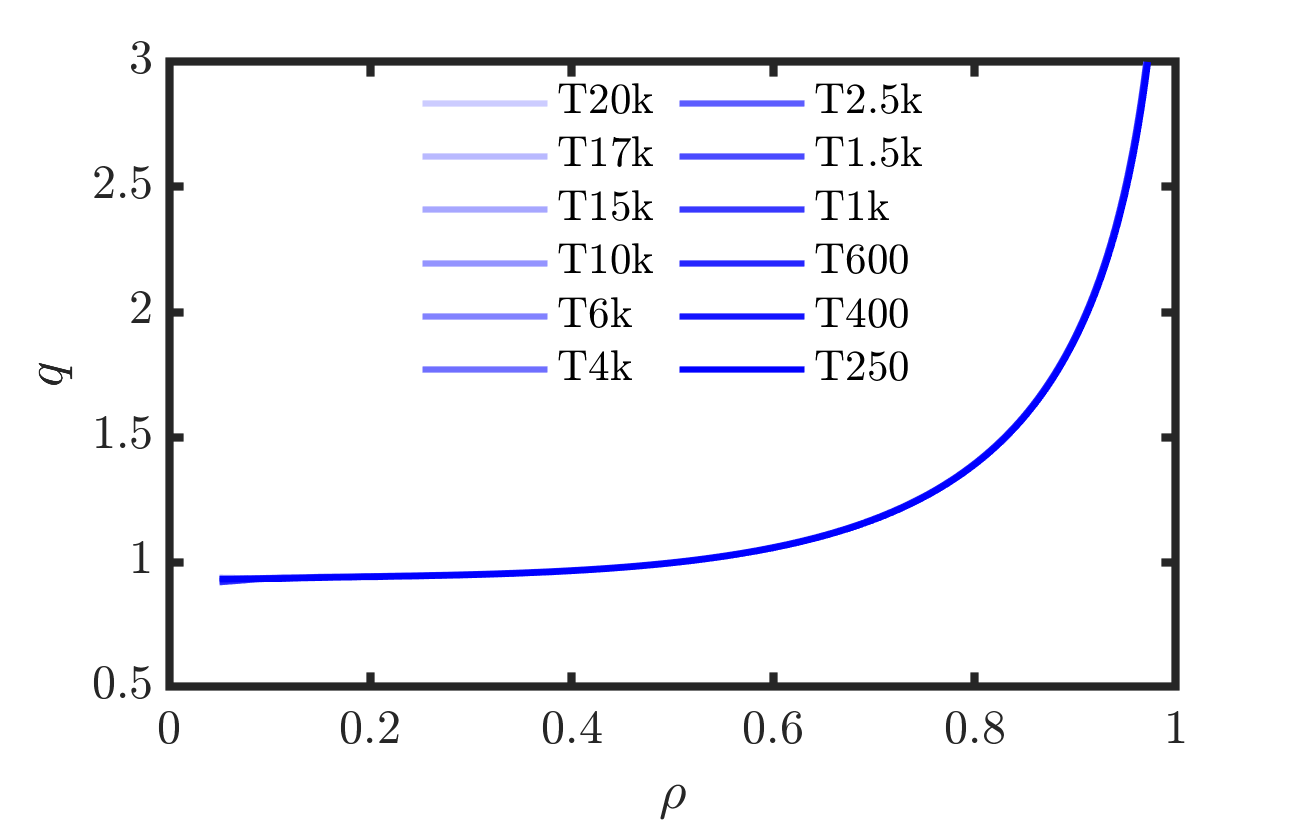}
      (a) $q$ profiles
      \end{minipage}
      
      \begin{minipage}[t]{0.45\textwidth}\centering
      \includegraphics[width=\textwidth]{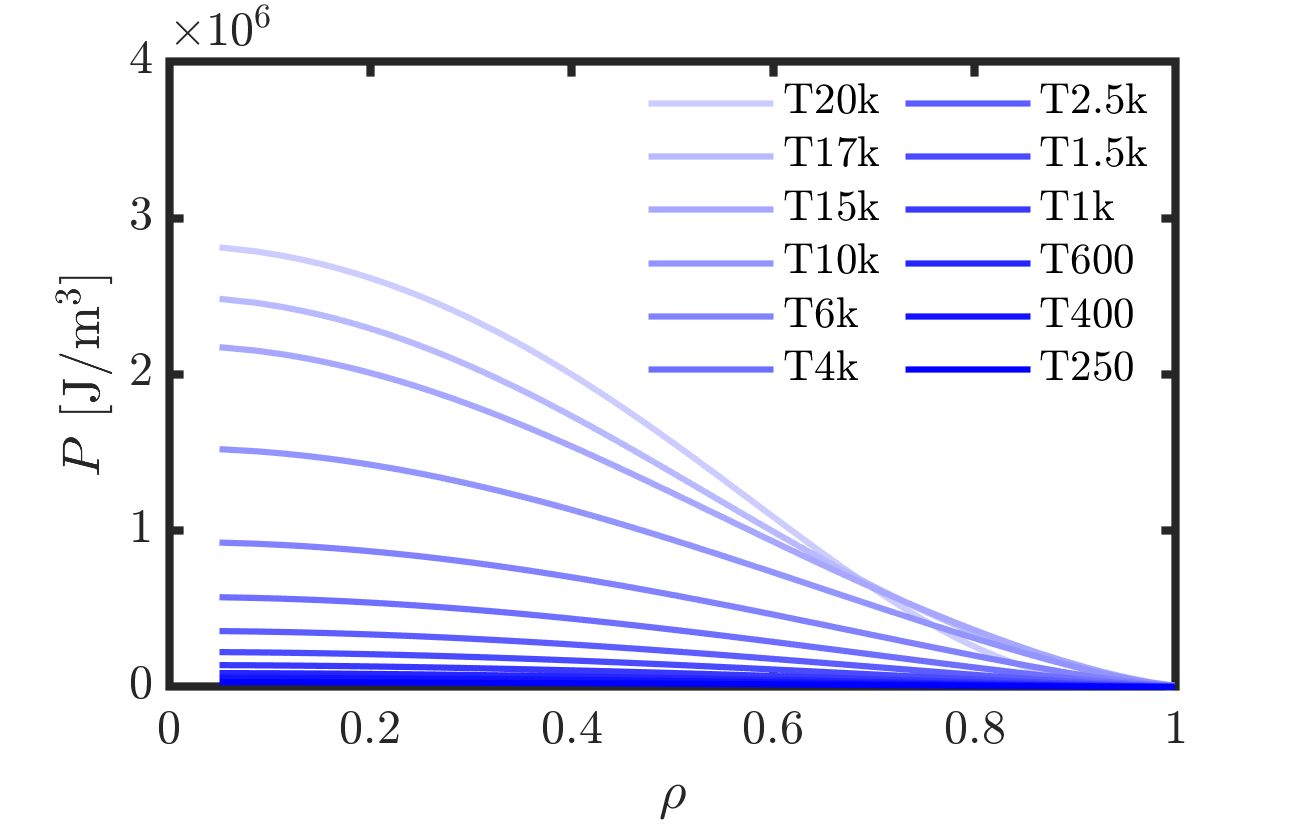}
      (b) $P$ profiles
      \end{minipage}
   \end{tabular}
   \caption{Comparison of $q$ profiles and $P$ profiles used in plasma $\beta$ scan. The scan is performed by relaxing temperature profiles with on-axis temperature in the range of $240$~eV -- $20$~keV ($\beta_0 = 0.052 - 4.75$\%). The label indicates the on-axis electron temperature for the given case. For example, ``T20k" means on-axis $T_{e,0}=20$~keV. The temperature profile drops from 20~keV to 240~eV while $q$ profiles overlap almost exactly as a single line.}
   \label{fig_beta_scan_profiles}
\end{figure}

The benefit of this method is to obtain different pressure profiles while keeping the density profile and the $q$ profile similar. The $q$ profiles and the pressure profiles for each simulated case are provided in Fig.~\ref{fig_beta_scan_profiles}. The profiles are labeled by their on-axis temperature, e.g., ``T20k" means the $T_0=20$~keV. It is clear that the $q$ profile remains unchanged while pressure gradients vary.

\begin{figure}[t]
   \centering
      \begin{minipage}[t]{0.6\textwidth}\centering
      \includegraphics[width=\textwidth]{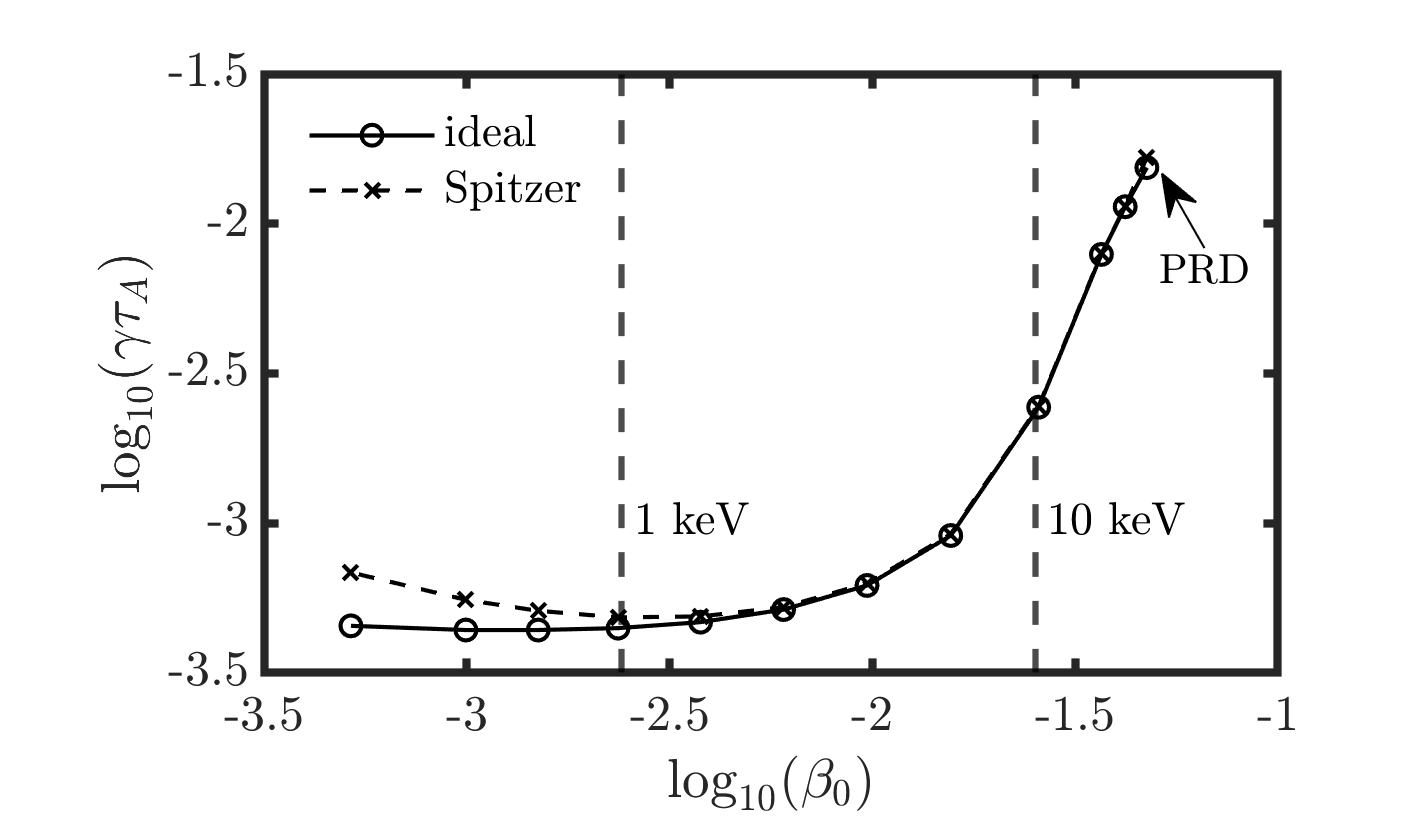}
      \end{minipage}
   \caption{Plasma $\beta$ scan by relaxing electron temperature profiles with on-axis temperature in the range of $240$~eV -- $20$~keV ($\beta_0 = 0.052 - 4.75$\%). The corresponding $q$ and $P$ profiles are provided in Fig.~\ref{fig_beta_scan_profiles}.
   Linear growth rate $\gamma$ is normalized by $\tau_A = 2.10\times 10^{-7}$ s. The baseline case is labeled at the right end of all data points, with $\beta_0 = 4.75$\%. The vertical lines show the simulation data points with $\beta_0 = 0.25$\% ($T_{e,0} \sim 1$~keV) and $\beta_0 = 2.5$\% ($T_{e,0} \sim 10$~keV) respectively. The ``x" data refers to simulations with Spitzer resistivity (with a dashed trending line), while ``o" data is for simulations with $\eta = 0$ (with a solid trending line). The whole device mesh is used in simulations.}
   \label{fig_beta_scan}
\end{figure}
\par
Fig.~\ref{fig_beta_scan} shows the plasma $\beta$ scan: a strong influence of pressure profile is observed when $T_{e,0}$ exceeds keV level. When $\beta_0$ exceeds 2\% (roughly $T_{e,0}>8.1$~keV), the growth rate increases rapidly by more than one order of magnitude. When $\beta_0 \rightarrow 0$, we begin to see the resistive effects destabilizing the mode. Simulations with the Spitzer resistivity (dashed line labeled ``Spitzer") give higher growth rates than the simulations by setting $\eta=0$ (solid line labeled ``ideal"). This is because a lower temperature leads to a higher collisional frequency, which makes the resistive effects more important.

\begin{figure}[t]
   \centering
      \begin{minipage}[t]{0.6\textwidth}\centering
      \includegraphics[width=\textwidth]{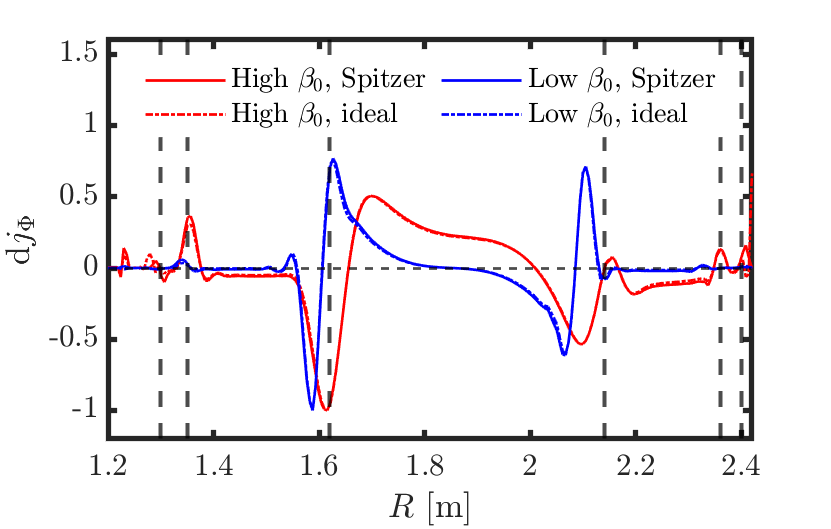}
      \end{minipage}
   \caption{Radial mode structures of perturbed toroidal current with the high $\beta_0=4.75$\%, with (labeled ``High $\beta_0$, Spitzer") and without Spitzer resistivity (labeled ``High $\beta_0$, ideal"); the low $\beta_0=0.052$\%, with (labeled ``Low $\beta_0$, Spitzer") and without Spitzer resistivity (labeled ``Low $\beta_0$, ideal"). The horizontal dashed lines indicate the zero location for mode amplitude reference. The vertical dashed lines denote the radial locations of the $q=1,2,3$ flux surfaces.}
   \label{fig_radial_mode}
\end{figure}

      

\par
The mode structures are also affected by different pressure profiles. We plot the radial mode profiles at $\Phi=0$, $Z=0$ for the four limiting cases in our parameter scan in Fig.~\ref{fig_radial_mode}. We separate the resistivity effects from the pressure effects by artificially setting $\eta$ to zero or Spitzer resistivity. Both high and low $\beta_0$ cases showed that the mode structure does not differ significantly with Spitzer resistivity or setting $\eta=0$. However, if we compare the high $\beta_0$ cases with the low $\beta_0$ cases, the pressure profile with a high temperature always gives a broader radial mode structure independent of the resistivity. This suggests that the $n=1$ mode is strengthened due to the pressure effects associated with the SPARC baseline equilibrium. The low-$\beta_0$ mode structure also shifts inwardly a little bit. This is likely due to the Shafranov shift when changing from a high pressure profile to a low pressure profile. For reference, the magnetic axis location in high-$\beta_0$ cases is at $R=1.89$~m, while the magnetic axis in low-$\beta_0$ cases is at $R=1.85$~m.

\subsection{\texorpdfstring{Plasma $\beta$}{plasma beta} effects on kink mode in SPARC using a 1D theoretical model}\label{subsec:beta_model_1d}

\par
To provide an intuitive understanding of the mode features we observe in the \simcode{} results, we investigate a simplified 1D radial model for a screw pinch geometry. Although there are geometric differences to compare with the SPARC results, the key physics is surprisingly well captured by this simple model as we will show in this section. For elucidating the underlying mechanisms of $n=1$ modes in SPARC, this model is sufficient. However, the results should not be interpreted as quantitative predictions for SPARC. A detailed discussion of such a simplified model is well summarized in the thesis work by Brochard, 2019\cite{Brochard2019}. In the following discussion, we will explain how the pressure drive and the current drive work together to affect the kink mode growth for the SPARC baseline equilibrium.
\par
Starting from the energy principle, Freidberg, 2014\cite{Freidberg2014} derived an expression of plasma energy for a general screw pinch geometry. By assuming a perturbed displacement of the form, $\bmvec{\xi}(\bmvec{r}) = \bmvec{\xi}(r)\exp(im\theta+ik_z z)$, one can simplify the potential energy (for internal modes, assume $\xi(a)=0$ at the minor radius $a$) and obtain an eigenvalue problem,

\begin{equation}
\label{eq_1d_screw_pinch}
\frac{\mathrm{d}}{\mathrm{d} r} \left[ \left(\rho\omega^2-f(r)\right)\frac{\mathrm{d}}{\mathrm{d} r}\xi(r) \right]+ g(r)\xi(r) = 0,
\end{equation}

\begin{equation}
\label{eq_f_term}
\mathrm{with \quad} 
f(r) = \frac{r \epsilon^2 B_t^2}{1+\epsilon^2 n^2/m^2} \left(\frac{1}{q}-\frac{n}{m}\right)^2,
\tag{\ref{eq_1d_screw_pinch}$.1$}
\end{equation}

\begin{equation}
\label{eq_g_term}
g(r) = \frac{\epsilon^2 B_t^2}{1+\epsilon^2 n^2/m^2} \left[ \frac{2\mu_0 P^\prime}{B_t^2}\frac{n^2}{m^2} +
       \frac{1}{r} \left(\frac{1}{q}-\frac{n}{m}\right)^2 \left( m^2-1\right) + \mathcal{O}(\epsilon^2) \right]
\tag{\ref{eq_1d_screw_pinch}$.2$}
\end{equation}

Here, $\rho$ represents the plasma density in this 1D model. We express the wavenumbers, $k_\theta=\frac{m}{r}$ and $k_z=-\frac{n}{R}$. The ratio, $\epsilon =\frac{r}{R}$, between the minor radius and the major radius is used as an ordering parameter, and we explicitly write out the leading terms in $\epsilon$. During the derivation, we make the assumption, $\xi \ll r_1 \xi^\prime$, to capture the internal kink mode feature in the inertia layer, where $q(r_1) = 1$ is the resonance surface for the 1/1 mode. In this way, the displacement that minimizes the energy can be approximated as $(\xi,\eta,\xi_\parallel) \approx (\xi,ir\frac{B_\theta}{B}\xi^\prime, ir\frac{B_t}{B} \xi^\prime)$. This will lead to an expression for the kinetic energy in terms of $\xi^\prime$, $K \approx 2\pi^2 R_0 \int_0^a dr \left( \rho\omega^2 r^3\xi^{\prime2} \right)$. The mode growth rate is therefore determined by $K(\omega^2) = -\delta W$, with the plasma energy given by $\delta W = 2\pi^2R_0/\mu_0\int_0^a dr\left(f\xi^{\prime 2}+g\xi^2\right)$. 
\par
By assuming the aspect ratio $\epsilon_a = a/R_0 \ll 1$, we can further simplify the expression for $m=1$, $n=1$ mode, noticing that $f(r) = r^3F^2(r)$ with the definition of $F(r) = \bmvec{k}\cdot\bmvec{B} = \frac{B_\theta}{r}\left(1-q\right)$, and the remaining pressure drive becomes $g(r) = \frac{q^2B_\theta^2}{B_t^2} 2\mu_0 P^\prime$. This lowest order approximation can be used to study the qualitative property of an $m=1$, $n=1$ internal kink mode in a large aspect ratio tokamak. A similar derivation starting from the reduced-MHD vorticity equation is used in an earlier work by Brennan, et al. 2014\cite{Brennan2014}, to investigate the 2/1 and 1/1 modes with a weakly reversed shear $q$ profile for DIII-D.
\par
\begin{figure}[t]
   \centering
      \begin{minipage}[t]{0.6\textwidth}\centering
      \includegraphics[width=\textwidth]{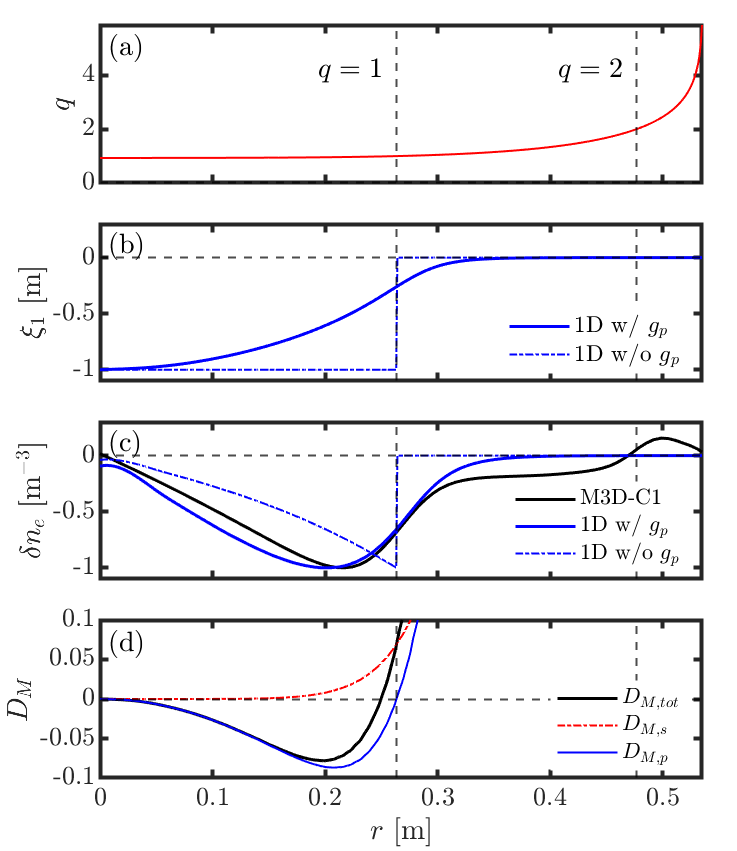}
      \end{minipage}
   \caption{Comparison of the radial mode structure from \simcode{} simulation and the $m=1$, $n=1$ eigenfunction, $\xi_{1,1}$, from a 1D simplified screw pinch model with a circular cross section. Panel (a) shows the ``q" profile as a function of the minor radius, $r = (R-R_0)\vert_{\Phi=0,Z=0}$. The eigenfunctions $\xi_{1,1}$ with (``1D w/ $g(r)$") and without pressure drive (``1D w/o $g(r)$") are provided in panel (b). The density perturbation $\delta n_e$ calculated from this 1D model is also compared with a radial output from \simcode{} (at $\Phi = 0$, $Z = 0$) in the panel (c). Panel (d) shows the calculated Mercier criterion based on 1D profiles. The $q=1$ and $q=2$ locations are labeled with vertical lines.}
   \label{fig_circular}
\end{figure}

To compare the 1D model and the \simcode{} simulations, we start from Eq.~\ref{eq_1d_screw_pinch} and use flux-averaged profiles, extracted from SPARC baseline equilibrium, as inputs to compute the coefficient functions $f(r)$ and $g(r)$. The toroidal field is $B_t = I/R$, where $I$ is the current function. The poloidal field is approximated as $B_\theta=r B_t/(R q)$. The radial coordinate is chosen based on a 1-1 mapping from the poloidal flux to a 1D radial cut at $\Phi=0$, $Z=0$ (from the magnetic axis to the LCFS). As an example, the converted $q$ profile is plotted in Fig.~\ref{fig_circular}a as a function of the radial location, $r \in (0,a) = (0,0.53)$. The $q=1$ location $r_1=0.26$ m is roughly in the center of the simulation domain.

\par
The original eigenvalue equation is first discretized by using a $C^0$-continuous finite element method (FEM), and then numerically solved by an eigenvalue solver in the SciPy package. The 1D eigenfunction $\xi_{1,1}$ for $m=1$, $n=1$ is plotted in Fig.~\ref{fig_circular}b, where subscript ``1,1" emphasizes the poloidal and toroidal mode numbers, $(m,n)=(1,1)$. We tested two cases, with and without the $g(r)$ function in Eq.~\ref{eq_1d_screw_pinch}. Without $g(r)$, an ideal kink mode solution will be obtained with a constant $\xi_{1,1}=C$ within $q=1$ flux surface (labeled ``1D w/o $g(r)$"). By including the $g(r)$, a pressure drive proportional to $P^\prime$ will take effect and smooth out the step function at $q=1$ (labeled ``1D w/ $g(r)$"). 
\par
The changes in $\xi_{1,1}$ are mainly due to the inclusion of pressure drive. This is because for $m=1$, the leading order term from the current in $g(r)$ vanishes. We can convert the displacement to the density perturbation $\delta n_e = -\bmvec{\xi}\cdot\nabla n_e = -\xi n_e^\prime(r)$, where the incompressibility $\nabla\cdot\bmvec{\xi} = 0$ is assumed for the most unstable mode. The computed perturbation is compared with the \simcode{} simulation directly in Fig.~\ref{fig_circular}c. With pressure drive $g(r)$, the peak density perturbation around $q=1$ surface is well captured by the 1D model, including the extended radial mode structure away from the resonance surface. The mode amplitude beyond $r=0.3$ m should be dominated by higher $m$ harmonics, since $\xi_{m,n}\to 0$ when $q>m/n$.
\par
For a quick assessment of the MHD stability, the Mercier criterion is also evaluated using a low-$\beta_0$, large-aspect-ratio approximation:

\begin{equation}
\label{eq_mercier}
D_{M,tot} = D_{M,s}+D_{M,P} = \left( \frac{r}{q} \frac{\mathrm{d}q}{\mathrm{d} r} \right)^2 + \frac{8\mu_0}{B_t^2} r \frac{\mathrm{d}P}{\mathrm{d} r} (1-q^2) \quad \text{$>0$ for stability}
\end{equation}
where $D_{M,s}$ represents the stabilizing effects from the magnetic shear and $D_{M,P}$ represents the destabilizing effects from strong pressure gradients. The calculated $D_{M,tot}$ is shown in Fig.~\ref{fig_circular}d. A negative contribution appears inside $q=1$ surface with a low magnetic shear and a large pressure gradient, indicating a strong pressure-driven destabilization of the MHD instability. This is also consistent with the perturbation analysis, that including $g(r)$ term greatly modifies the mode structure.
\par
\begin{figure}[t]
   \centering
   \begin{tabular}{cc}
      \begin{minipage}[t]{0.45\textwidth}\centering
      \includegraphics[width=\textwidth]{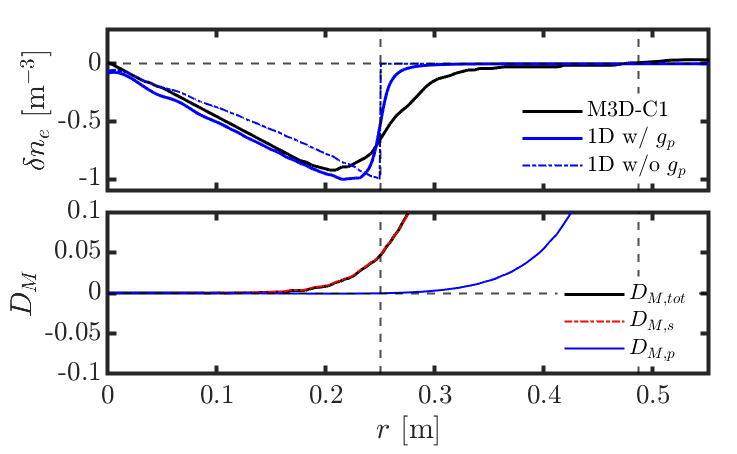}
      (a) $q_0=0.93$, $\beta_0=0.052$\%
      \end{minipage}
      
      \begin{minipage}[t]{0.45\textwidth}\centering
      \includegraphics[width=\textwidth]{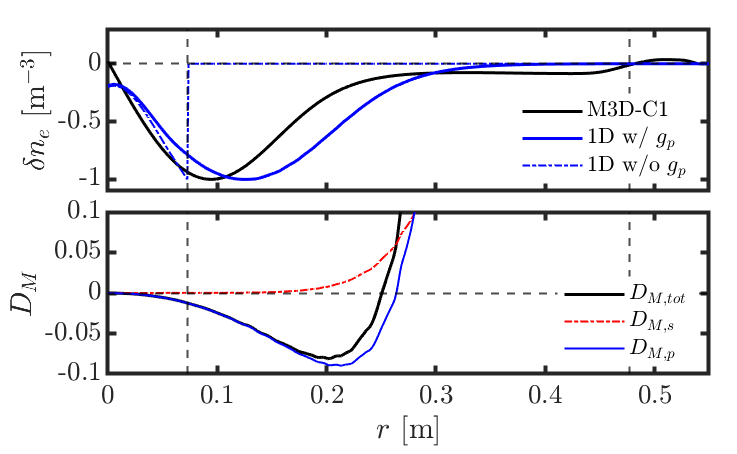}
      (b) $q_0=0.991$, $\beta_0=4.75$\%
      \end{minipage}
   \end{tabular}
   \caption{Comparison of the radial mode structures from \simcode{} simulation and the $m=1$, $n=1$ eigenfunction, $\xi_{1,1}$, from a 1D simplified screw pinch model. The simulated two cases separate the current drive and the pressure drive to show their effects on the mode structures and the Mercier criteria.}
   \label{fig_circular_compare}
\end{figure}

To further distinguish the current-driven and pressure-driven contributions, similar calculations of $\delta n_e$ and $D_{M,tot}$ are performed for two additional cases: $q_0=0.93$, $\beta_0=0.052$\% and $q_0=0.991$, $\beta_0=4.75$\% as shown in Fig.~\ref{fig_circular_compare}. As expected, the low-$\beta_0$ case shows a localized kink mode structure with a negligible contribution of $D_{M,tot}$ (Fig.~\ref{fig_circular_compare}a), whereas the high-$\beta_0$ case shows an extended mode structure beyond $q=1$ location, associated with the strong pressure gradient (Fig.~\ref{fig_circular_compare}b). These results support the above discussion that both current drive and pressure drive contribute significantly to the instability in the SPARC baseline case.
\par
The solved mode growth rate is also qualitatively comparable. For $g(r)=0$ case, the 1/1 mode is marginally stable with a very low frequency $\omega\tau_A = 1.36\times10^{-4}$. Including the higher order current terms in $g(r)$ destabilizes the mode to give a growth rate of $\gamma\tau_A = 4.26\times10^{-3}$. Further including the pressure term in $g(r)$, a growth rate of $\gamma\tau_A = 5.68\times10^{-2}$ is obtained which is 3--4 times the \simcode{} simulation value. The drastic increase after including the pressure gradient term is consistent with the linear scan in Fig.~\ref{fig_beta_scan}. When we remove pressure evolution in the \simcode{} simulation, the growth rate drops from $\gamma\tau_A = 1.67\times10^{-2}$ to $\gamma\tau_A = 7.45\times10^{-4}$ (labeled as ``PRD ($\beta_0=0$)'').

\par
Moreover, a drastic decrease of the growth rate is also reproduced with the 1D model, when we use the low $\beta_0 = 0.052$\% equilibrium as in Fig.~\ref{fig_beta_scan}. The growth rate using 1D model is $\gamma\tau_A=5.24\times10^{-3}$, while for \simcode{}, $\gamma\tau_A = 4.5\times10^{-4}$ without resistivity. Though not presented in figures, the higher order current drive $\mathcal{O}(\epsilon^2)$ in $g(r)$ became more important in determining the radial mode structure, when the pressure drive gets weaker at low $\beta_0$, and the radial mode also qualitatively matches the \simcode{} simulation after including the current terms.
\par
Apart from 1/1 mode results, the 1D model also provides insights on mode structures with higher poloidal harmonics. For SPARC baseline case, the inclusion of $P^\prime$ term can also be used to solve higher $m$ harmonics for $m=1\sim 5$, which can be responsible for the bump structure beyond $q=1$ surface. The mode becomes more and more stable as $m$ number increases when we input different $m$'s into the 1D solver. This can explain the features in Fig.~\ref{fig_ntor_scan}b: for the $n=1$ mode, the $m$-th component will dominate between $q=m-1$ and $q=m$ flux surfaces, as the $(m-1)$-th component vanishes around $q=m-1$ while the $(m+1)$-th component is more stable compared to the $m$ component. 
\par
A scaling trend of roughly $\gamma\tau_A\propto n^{-2.1}$ can be observed in Fig.~\ref{fig_ntor_scan}a. A partial explanation for this trend can be obtained with the help of a 1D cylindrical model. For a fixed m/n ratio, most terms in $f(r)$ and $g(r)$ are identical so they won't result in a scaling trend for any $m=n$ modes. However, a weak scaling of the growth rate for the $m=n$ mode can exist when including the $(m^2-1)$ term in $g(r)$. This 1D cylindrical approximation only leads to $\gamma\tau_A\propto n^{-0.9}$ (for $m=n$), instead of the observed $n^{-2.1}$ in \simcode{} simulations (Fig.~\ref{fig_ntor_scan}a). This suggests that a more complete model considering the complex geometry may be needed to explain this trend. That is, a global linear stability code is required for a more accurate comparison with the \simcode{} simulations for the SPARC.
Such stability code studies are beyond the scope of this paper and may be considered for benchmark purposes in future works.


\section{Nonlinear simulation of initial sawtooth crash for SPARC baseline-like equilibria}\label{sec:nonlinear}


\par
In the linear simulations, we have identified the most unstable mode to be an $m=1$, $n=1$ internal kink mode. In this section, we discuss the saturation of the kink mode and the profile changes during the nonlinear stage. An interpretation of the observed sawtooth crash is provided including consistent features from both the Kadomtsev model and the Wesson model. These models are widely used to explain previous sawtooth measurements in various devices. 
\par
As was shown in linear parameter scans, both the plasma $\beta$ and $q$ profile can affect the 1/1 mode in SPARC significantly. To systematically evaluate sawtooth crashes in SPARC, we separate the pressure drive and the current drive by modifying the profiles based on the SPARC baseline equilibrium. Firstly, we focus on the marginally unstable case, as shown in our linear MHD simulations, with $q_0$ approaching unity. In these cases, the pressure drive is kept comparable to the PRD design point, with a fixed $\beta_0=4.75$\%. Next, we lower the temperature profile while fixing the current drive unchanged, with an on-axis $q_0$ around 0.93. Eventually, the sawtooth crash in the baseline case is presented and the mechanism of the sawtooth crash is provided in Sec.~\ref{subsec:sawtooth_model}. Our simulation results provide insights on the onset of sawtooth crash in SPARC and serve as a first step to deliver a reliable sawtooth modeling of MHD events in SPARC tokamak.

\subsection{Influence of the wall boundary condition and the toroidal mode resolution}\label{subsec:numer_converge}

\par
\begin{figure}[t]
   \centering
   \begin{tabular}{cc}
      \begin{minipage}[t]{0.48\textwidth}\centering
      \includegraphics[width=\textwidth]{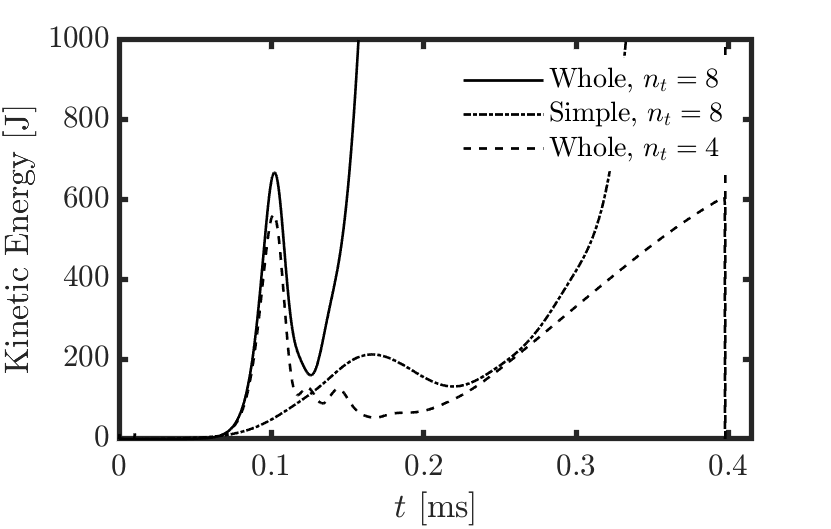}
      (a) time history of kinetic energy
      \end{minipage}

      \begin{minipage}[t]{0.48\textwidth}\centering
      \includegraphics[width=\textwidth]{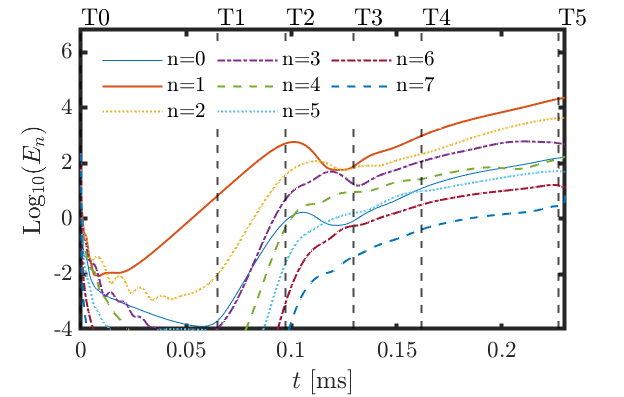}
      (b) kinetic energy spectrum with the whole mesh geometry and $n_t=8$
      \end{minipage}
\\
      \begin{minipage}[t]{0.48\textwidth}\centering
      \includegraphics[width=\textwidth]{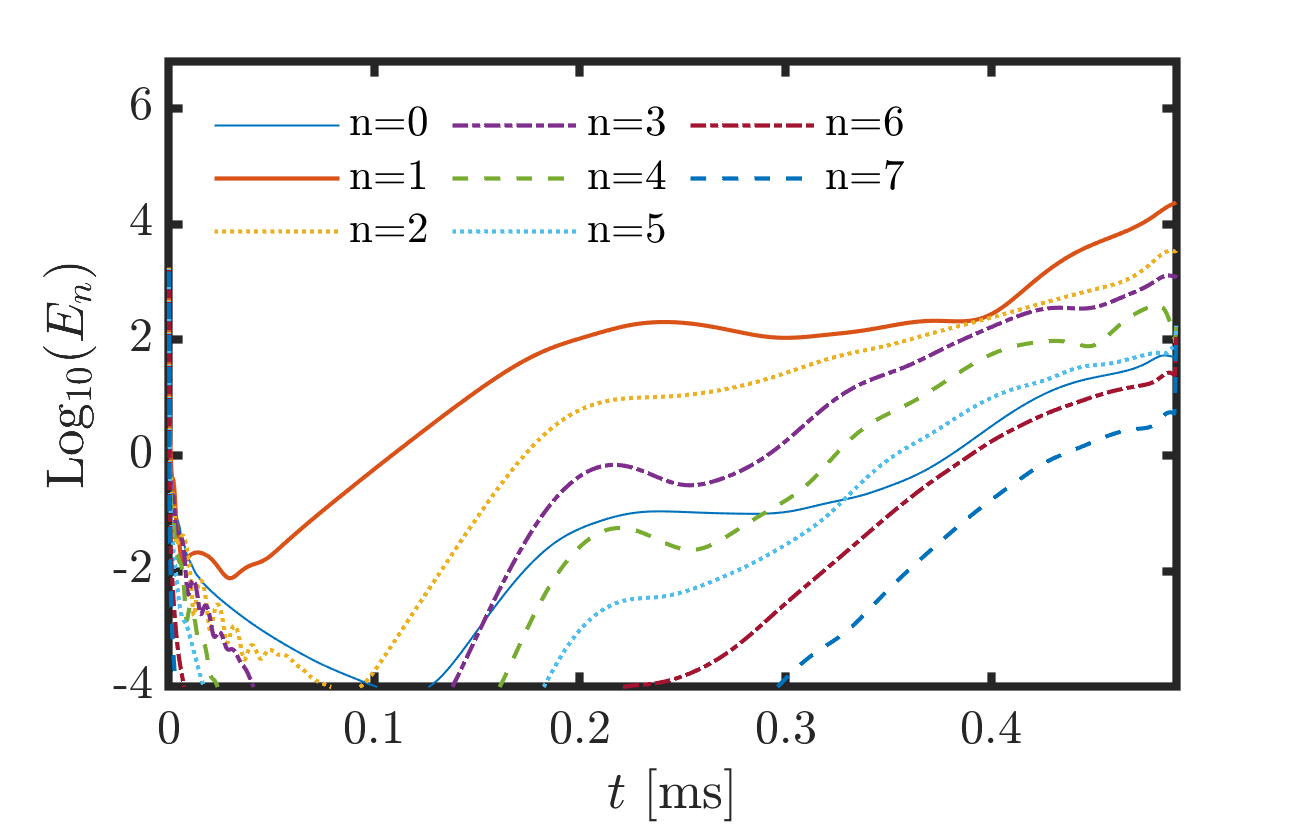}
      (c) kinetic energy spectrum with the simple mesh geometry and $n_t=8$
      \end{minipage}

      \begin{minipage}[t]{0.48\textwidth}\centering
      \includegraphics[width=\textwidth]{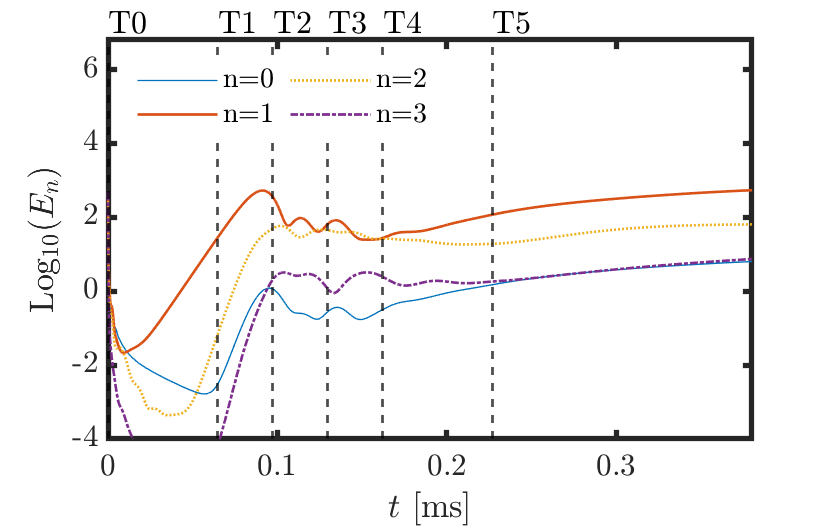}
      (d) kinetic energy spectrum with the whole mesh geometry and $n_t=4$
      \end{minipage}
      
   \end{tabular}
   \caption{Panel (a) shows the comparison of mode kinetic energies with toroidal resolutions of $n_t=4$ and $n_t=8$, and also compares the effects of switching the whole device mesh grids to the simple mesh. Panel (b) is the time history of kinetic energy spectrum with the whole device mesh and $n_t=8$. The toroidal mode resolution is up to $n=7$. The dashed lines mark six time slices (T0=0~$\mu$s, T1=65~$\mu$s, T2=97~$\mu$s, T3=130~$\mu$s, T4=162~$\mu$s and T5=227~$\mu$s). Panel (c) shows the kinetic energy spectrum with the simple mesh and $n_t=8$. Panel (d) shows the whole device mesh and $n_t=4$. The toroidal mode resolution is up to $n=3$ and the dashed lines in Panel~d mark the same time slices T0--T5 as in Panel~b.}
   \label{fig_nonli_numer_compare}
\end{figure}

To assess the robustness of nonlinear results, we perform simulations with different toroidal resolutions and mesh geometries. The corresponding time evolutions of the kinetic energy are shown in Fig.~\ref{fig_nonli_numer_compare}. For the whole device simulation, reducing the toroidal resolution from $n_t=8$ to $n_t=4$ still resolves the initial saturation phase of instabilities: the kinetic energy saturates at 667~J with $n_t=8$, compared to 562~J with $n_t=4$. Both exhibit comparable growth rates for the dominant $n=1$ component as in the linear simulation. 
\par
The two simulations deviate after $\sim t=0.13$~ms, which corresponds to T3 in Fig.~\ref{fig_nonli_numer_compare}b. The toroidal components of kinetic energy with $n_t=4$ are shown in Fig.~\ref{fig_nonli_numer_compare}d. The rapid growth in the second phase after T3 is greatly reduced with the lower toroidal resolution. With $n_t=8$, $n=2$ and $n=3$ components grow to comparable amplitudes to the $n=1$ component at T3, but $n=3$ component stays about one order of magnitude lower with $n_t=4$. This indicates that sufficient toroidal resolution is required for quantitatively resolving the long-time nonlinear evolution of high-$n$ components. For the present purpose of studying the initial sawtooth crash, $n_t=4$ provides an adequate description. However, higher toroidal resolution should be considered for quantitative predictions of the subsequent nonlinear sawtooth crash evolution.

\par
Regarding the influence of boundary conditions on the nonlinear evolution,  a simulation using the simple geometry with $n_t=8$ is performed (Fig.~\ref{fig_nonli_numer_compare}c). Consistent with the lower growth rate in linear analysis, the simple geometry exhibits a slower nonlinear evolution due to the close conducting boundary, leading to a lower initial saturation level of kinetic energy ($\sim 212$~J), as shown in Fig.~\ref{fig_nonli_numer_compare}a. The second phase of rapid growth with high-$n$ components is also reproduced with the simplified geometry. That is, the qualitative analysis of the initial saturation mechanisms presented in the subsequent parameter studies remains valid. However, the characteristic timescales can be modified after including resistive wall geometries in simulations.
\par
In Sec.~\ref{subsec:q_scan} and Sec.~\ref{subsec:kappa_scan}, we use a reduced set-up, a simple mesh geometry with $n_t=4$, for the modified baseline-like simulations. These simulations will give a slower mode growth due to the simple mesh geometry and the long term nonlinear evolution of high-$n$ components may be suppressed due to limited toroidal resolution. However, to study the initial sawtooth crash mechanism, these settings should be sufficient. The PRD baseline case will be solved with the highest resolution in this study, using the whole device geometry with $n_t=8$ to obtain the most accurate plasma evolution in the nonlinear stage.


\subsection{Effects of \texorpdfstring{$q$}{q} profile on the nonlinear sawtooth crash}\label{subsec:q_scan}

\par
\begin{figure}[t]
   \centering
   \begin{tabular}{cc}
   
      \begin{minipage}[t]{0.48\textwidth}\centering
      \includegraphics[width=\textwidth]{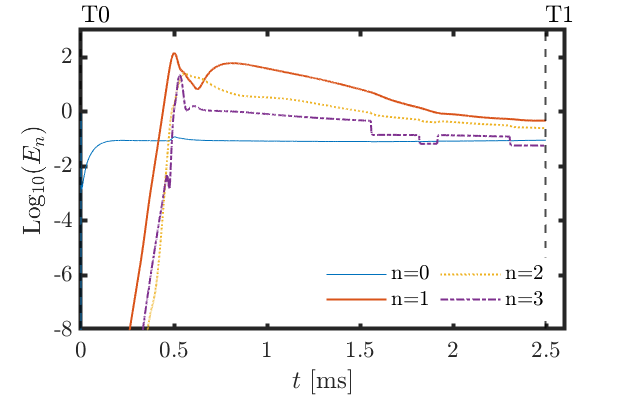}
      \end{minipage}
      
      \begin{minipage}[t]{0.48\textwidth}\centering
      \includegraphics[width=\textwidth]{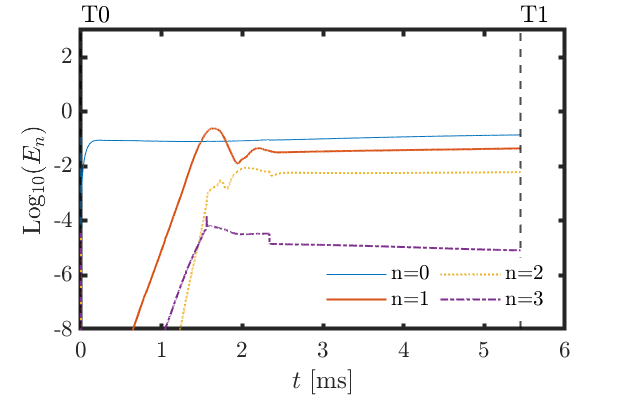}
      \end{minipage}

\\
      
      \begin{minipage}[t]{0.48\textwidth}\centering
      \includegraphics[width=\textwidth]{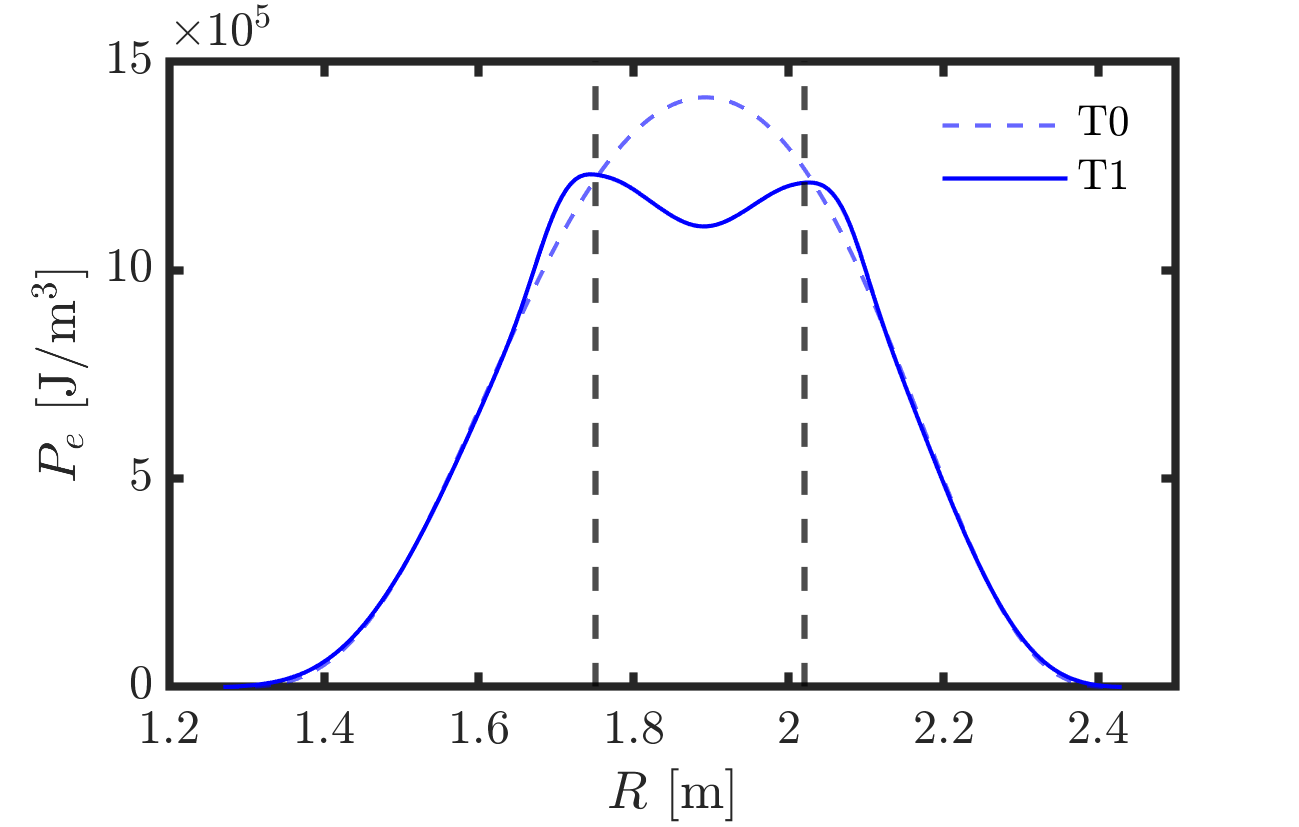}
      \end{minipage}
      
      \begin{minipage}[t]{0.48\textwidth}\centering
      \includegraphics[width=\textwidth]{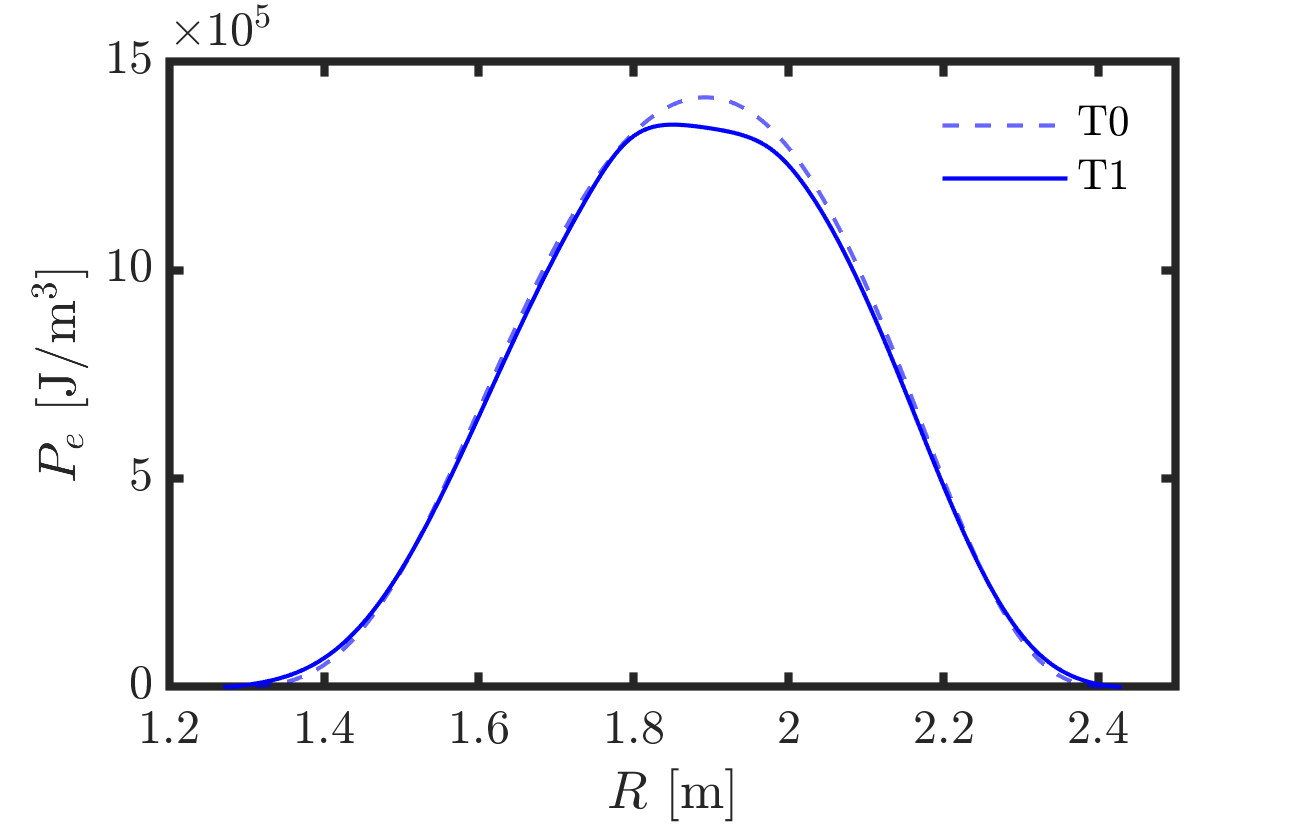}
      \end{minipage}
      
\\

      \begin{minipage}[t]{0.48\textwidth}\centering
      \includegraphics[width=\textwidth]{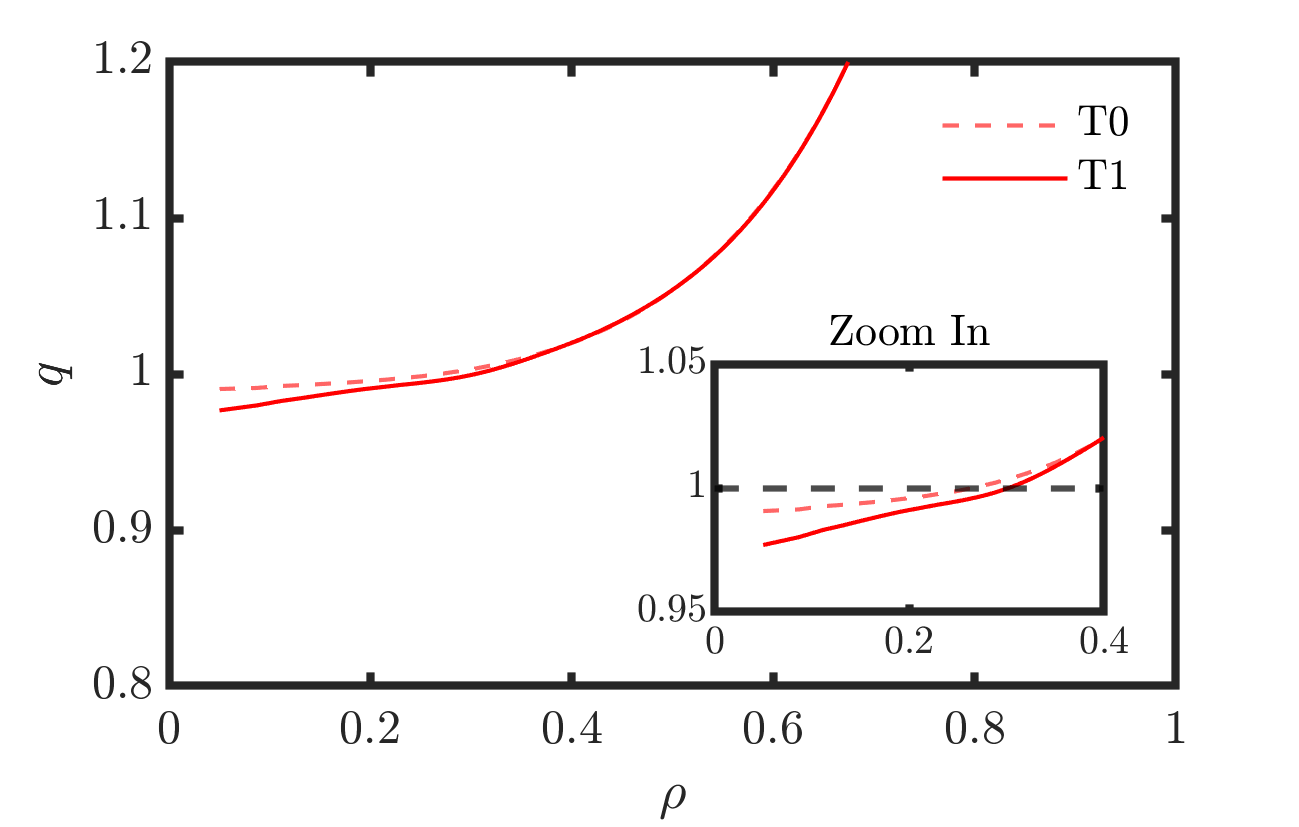}
      (a) $q_0 = 0.991$ ($F_s = 1.07$)
      \end{minipage}
      
      \begin{minipage}[t]{0.48\textwidth}\centering
      \includegraphics[width=\textwidth]{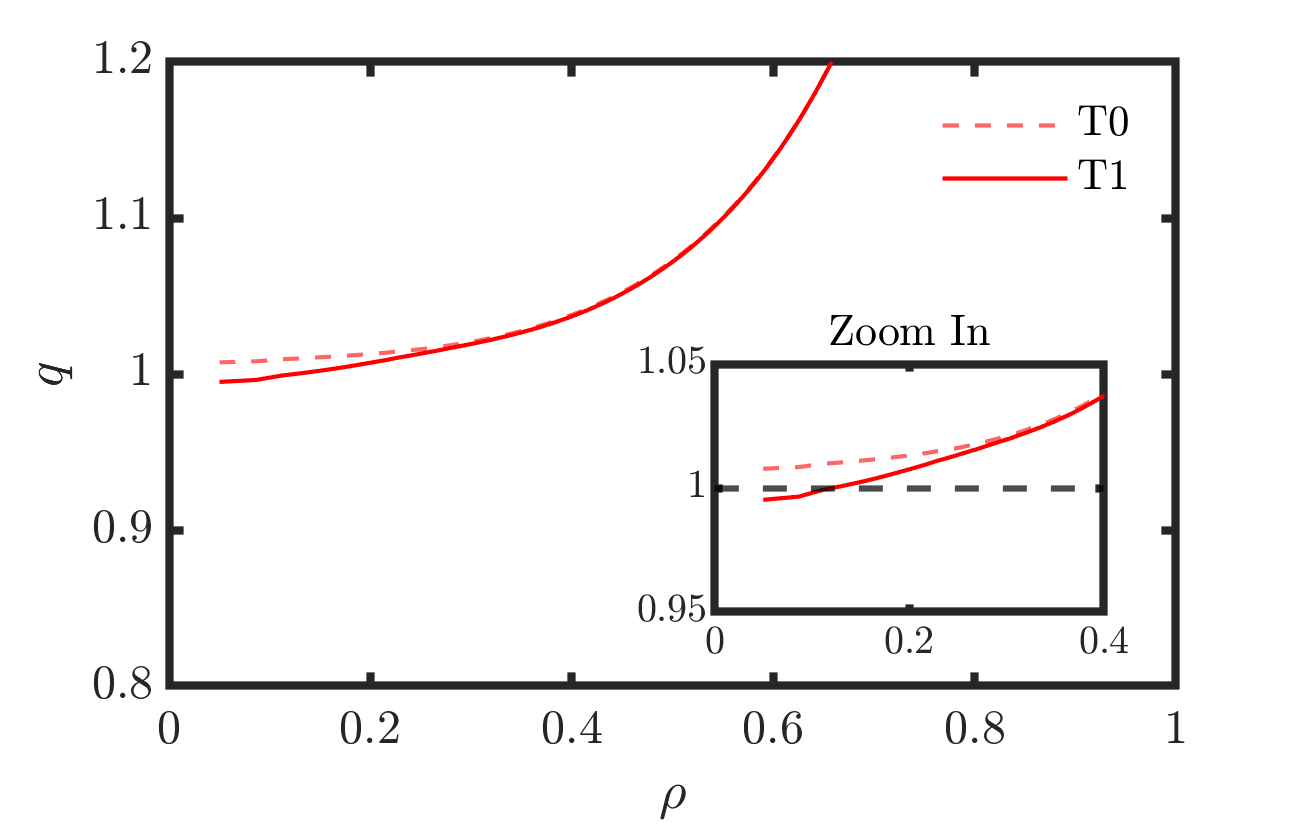}
      (b) $q_0 = 1.008$ ($F_s = 1.09$)
      \end{minipage}

   \end{tabular}
   \caption{Profile crashes after rescaling the equilibrium $q$ profile with a Bateman scaling factor, $F_s$. Column (a) is simulated with $q_0=0.991$ ($F_s = 1.07$). Column (b) is with $q_0=1.008$ ($F_s = 1.09$). The plotted quantities are time history of kinetic energy spectra up to $n=3$, electron pressure $P_e$ and safety factor $q$ from top to bottom. The 2 time slices feature the profiles before (T0) and after (T1) the profile crash of the nonlinear mode. The simple geometry mesh is used in these nonlinear simulations with toroidal resolution of $n_t=4$ planes.}
   \label{fig_q_shift}
\end{figure}

\par
To change the current drive while keeping the pressure drive unchanged, we can shift the $q$ profile to reduce the volume with $q<1$, so that the 1/1 mode becomes weaker. Two cases with $q_0=0.991$ ($F_s = 1.07$) and $q_0=1.008$ ($F_s = 1.09$) will be discussed separately in this section to study the effects of $q$ profile on the sawtooth crash in SPARC baseline-like scenarios. Following the numerical-sensitivity assessment in Sec.~\ref{subsec:numer_converge}, we use the simplified mesh grid with $n_t=4$ that includes only the plasma region, assuring a more efficient setup for the parametric study while retaining the essential physical mechanisms.
\par
In Fig.~\ref{fig_q_shift}, the time history of the kinetic energy spectrum is plotted for up to $n=3$. Although some high-$n$ components, e.g., $n=4$, can still be obtained numerically due to the higher order basis functions used in M3D-C1, they are not well-resolved with the limited toroidal resolution of $n_t=4$. The time history indicates an initial saturation of MHD instability growth at $t=0.5$~ms for the $q_0=0.991$ case, and at $t=1.6$~ms for the $q_0=1.008$ case. The pressure profiles before and after the mode saturation are plotted for both $q_0=0.991$ (Fig.~\ref{fig_q_shift}a) and $q_0=1.008$ (Fig.~\ref{fig_q_shift}b), while the initial pressure profiles are the same as those of the baseline case (with $\beta_0=4.75$\%). As the initial $q$ profile (before crash) gets closer to 1, the profile flattening after the sawtooth crash is reduced. 
\par
For $q_0=0.991$, although the $q<1$ volume is much smaller compared to the baseline case, a sawtooth crash is still presented near the magnetic axis with a hollowed pressure profile. The linear growth rate of the 1/1 mode is still as large as $\gamma\tau_A = 1.18\times 10^{-2}$, though the current drive is greatly reduced. (Note the baseline case with the simple mesh gives $\gamma\tau_A = 1.36\times 10^{-2}$.) Given the weak current drive with $q_0\sim 1$, the instability is mainly driven by the strong pressure drive associated with the high $\beta_0=4.75$\% ($20$~keV), which has already been confirmed in our linear analysis. The hollowed plasma profile brings our attention to the sawtooth crash observations in previous high-temperature tokamaks, such as the JET results.\cite{Granetz1988,Wesson1986}
\par
In the case of $q_0=1.008$, the instability quickly approaches the marginality. For simulation with the whole device mesh, the mode is marginally stable. For simulation with the simple mesh, a marginal linear growth rate of $\gamma\tau_A=2.48\times10^{-3}$ is observed. In the nonlinear simulation using the simple mesh, the profile is hardly flattened in this case. The rapid drop of the growth rate is also consistent with features of the quasi-interchange mode driven by the pressure effects with a flat $q$ profile. For the kink mode driven by the current drive, the instability will be reduced when $q_0$ rises far above 1. The strong pressure drive maintains the $n=1$ instability even when $q_0$ is slightly above 1. As we can see in the third row of Fig.~\ref{fig_q_shift}, the $q$ profile hardly changes when the pressure profile evolves considerably. This conclusion is further supported when we analyze the magnetic field and pressure evolution in the poloidal plane in the later subsections.


\subsection{Effects of pressure profile on the nonlinear sawtooth crash}\label{subsec:kappa_scan}

\par
\begin{figure}[t]
   \centering
   \begin{tabular}{cc}
      \begin{minipage}[t]{0.48\textwidth}\centering
      \includegraphics[width=\textwidth]{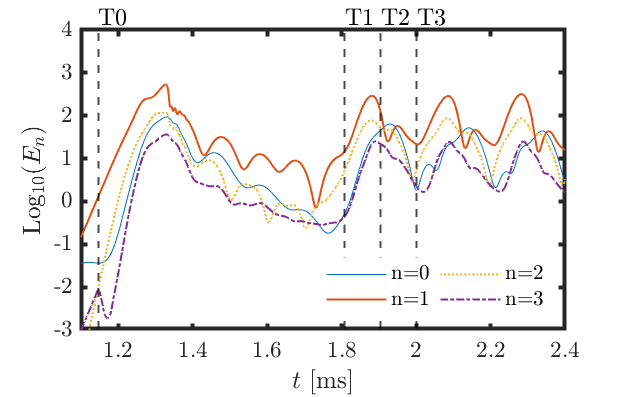}
      \end{minipage}
      
      \begin{minipage}[t]{0.48\textwidth}\centering
      \includegraphics[width=\textwidth]{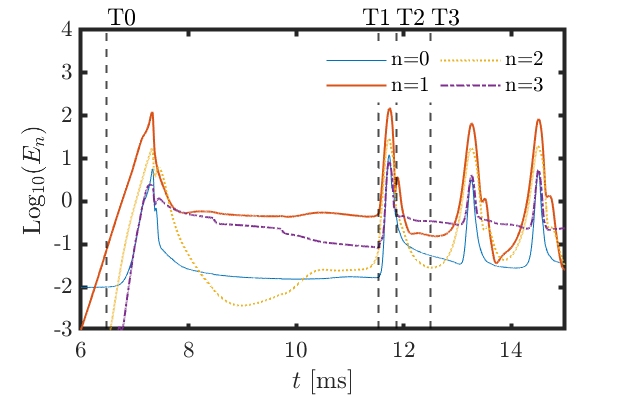}
      \end{minipage}
      
\\

      \begin{minipage}[t]{0.48\textwidth}\centering
      \includegraphics[width=\textwidth]{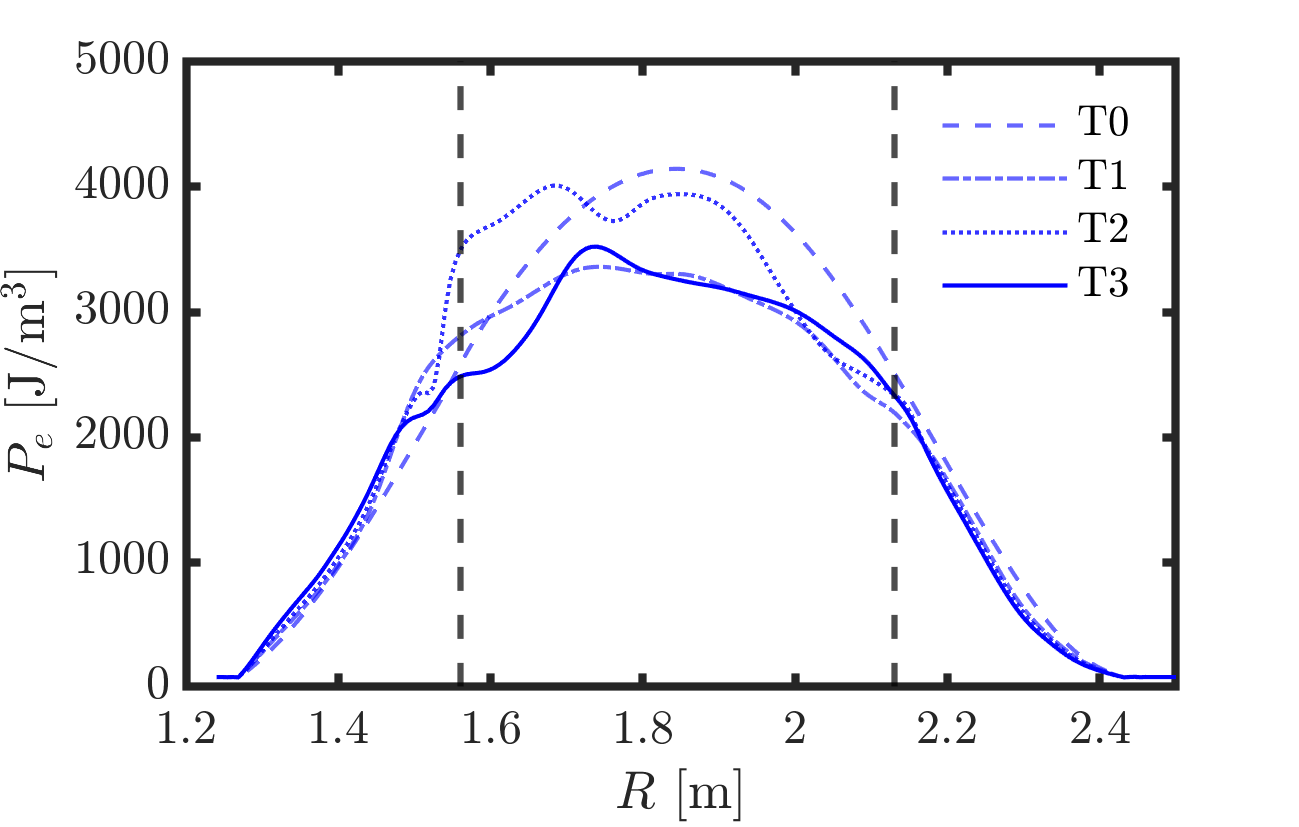}
      \end{minipage}
      
      \begin{minipage}[t]{0.48\textwidth}\centering
      \includegraphics[width=\textwidth]{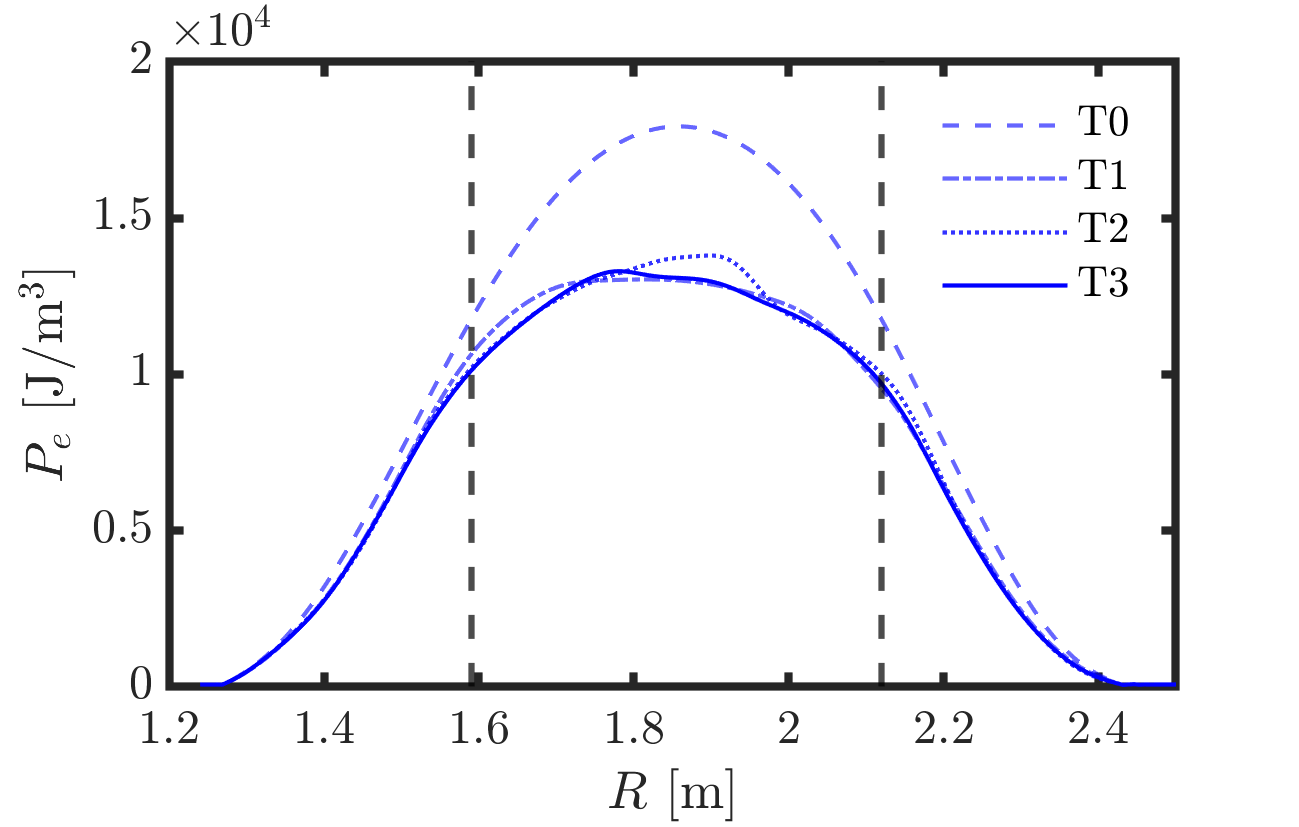}
      \end{minipage}

\\

      \begin{minipage}[t]{0.48\textwidth}\centering
      \includegraphics[width=\textwidth]{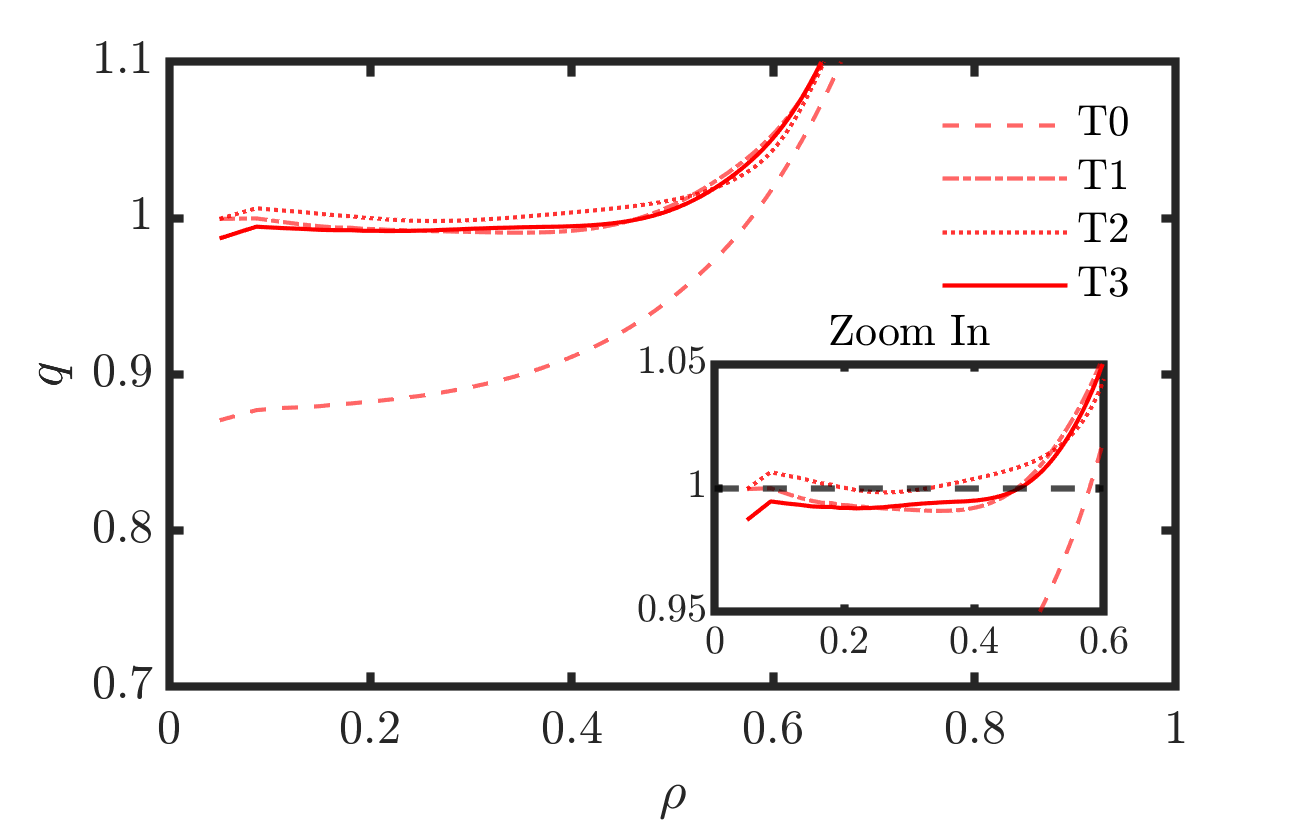}
      (a) $\beta_0 = 0.009$\% ($\bar{\kappa}_T=1.08\times 10^{5}$ m$^2$/s)
      \end{minipage}
      
      \begin{minipage}[t]{0.48\textwidth}\centering
      \includegraphics[width=\textwidth]{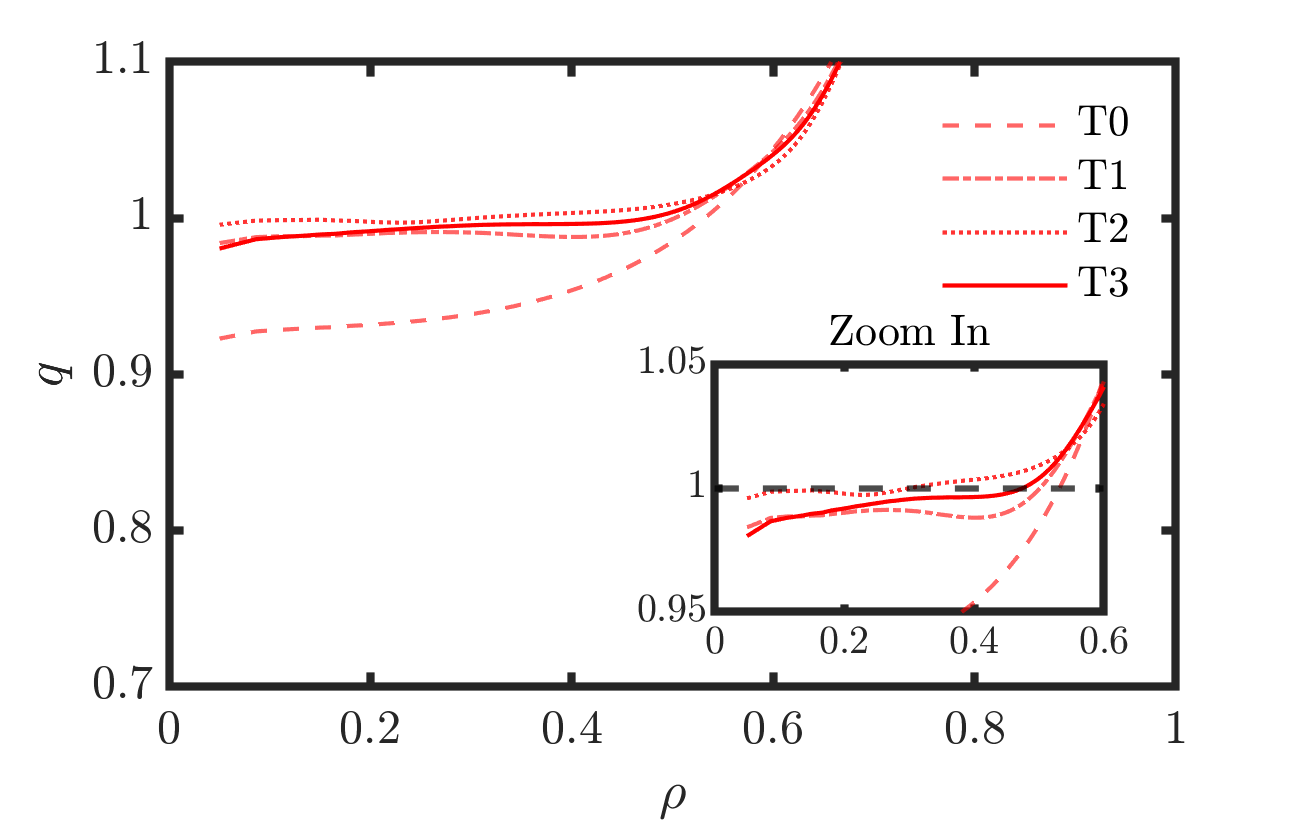}
      (b) $\beta_0 = 0.052$\% ($\bar{\kappa}_T=1.54 \times 10^{3}$ m$^2$/s)
      \end{minipage}
   \end{tabular}
   \caption{Sawtooth oscillations with low temperature profiles as the initial equilibria: Column (a) with heat conductivity $\bar{\kappa}_T=1.08\times 10^{5}$ m$^2$/s; Column (b) with heat conductivity $\bar{\kappa}_T=1.54\times 10^{3}$ m$^2$/s. The plotted quantities are time history of kinetic energy spectra, electron pressure $P_e$ and safety factor $q$ from top to bottom. The time history of kinetic energy spectra is resolved for up to $n=3$. The dashed lines mark four time slices (T0-T3) showing the initial equilibria (T0) and the profile before (T1), during (T2) and after (T3) a periodic sawtooth oscillation. The simple geometry mesh is used in these nonlinear simulations with toroidal resolution of $n_t=4$ planes.}
   \label{fig_kappa_scan}
\end{figure}

To modify the pressure profile, we increase the thermal heat conductivity to an extremely large value, similar to Sec.~\ref{subsec:beta_scan}. In Fig.~\ref{fig_kappa_scan}, we show the nonlinear results with $\bar{\kappa}_T = 1.08\times 10^{5}$ m$^2$/s and $\bar{\kappa}_T = 1.54\times 10^{3}$ m$^2$/s. For reference, the original SPARC baseline simulation uses a uniform perpendicular heat conductivity $\bar{\kappa}_T = 1.54$ m$^2$/s with $\beta_0 = 4.75$\%. 
\par
The temperature profile relaxes to a lower equilibrium value on the diffusive timescale controlled by perpendicular $\bar{\kappa}_T$: for $\bar{\kappa}_T = 1.08\times 10^{5}$ m$^2$/s, the new equilibrium is achieved after $\sim 1.5\times 10^{-2}$~ms with $\beta_0=0.009$\%; for $\bar{\kappa}_T = 1.54\times 10^{3}$ m$^2$/s, the equilibrium is achieved after $\sim 1.2$~ms with $\beta_0=0.052$\%. Based on these 2D equilibria, we continue the simulations with a 3D nonlinear setup. Multiple toroidal $n$ components are included with a dominant $n=1$ component in the early stage. Then, profile flattening processes occur with $q_0$ rising up above unity, suggesting a typical magnetic reconnection event in the sawtooth crash theory.\cite{Kadomtsev1975,Sykes1976}

\par
In Fig.~\ref{fig_kappa_scan}a, the time history of the kinetic energy spectrum for $\beta_0 = 0.009$\% case is provided. The plot begins from $t=1.2$ ms, by which time the system has already relaxed to the new equilibrium with a peaked electron temperature, $T_{e,0}=50$~eV. After the initial sawtooth crash, the system goes through a relatively long recovery period where the $q_0$ starts to drop slightly below 1 again and then evolves into periodic sawtooth crashes.
\par
As shown in Fig.~\ref{fig_kappa_scan}, a moderate sawtooth crash with 20-30\% flattening of the peaked profiles can be expected when we have low-$\beta_0$ profiles in SPARC. These sawtooth crashes are mainly driven by the internal kink mode with the low $q_0=0.93$. Moreover, unlike the pressure-driven cases in Sec.~\ref{subsec:q_scan}, no hollowed pressure structure is observed in these low-$\beta_0$ simulations. In the following subsection, we will move back to the baseline case combining both a low $q_0$ and high $\beta_0$, and study the baseline result based on the pressure-driven and current-driven sawtooth crash we just simulated.

\subsection{Initial sawtooth crash based on the SPARC baseline equilibrium}\label{subsec:prd_sawtooth}

In order to explore the underlying effects when both strong pressure and current drives are present, we analyze in this section the baseline case. As we have seen in previous discussion, the sawtooth crash can be driven by either the current drive (a low $q_0$) or the pressure drive (high plasma $\beta$). If there are additional stabilizing effects, such as in the TRANSP simulation where the heating and the alpha particles are taken into account, it is possible that the plasma evolves to a stage where both the current drive and the pressure drive become strong.\cite{Rodriguez-Fernandez2020} This leads to our next-step investigation of the baseline case, originating from the PRD design point, to study the sawtooth crash process when both the current, $q_0=0.93$, and the pressure, $\beta_0=4.75$\%, can destabilize the 1/1 mode in SPARC. As will be shown by the following results, a significant crash is observed, which makes this scenario unlikely to occur in practice unless sufficient stabilizing effects are included. 

\begin{figure}[h]
   \centering
   \begin{tabular}{cc}
   
      \begin{minipage}[t]{0.48\textwidth}\centering
      \includegraphics[width=\textwidth]{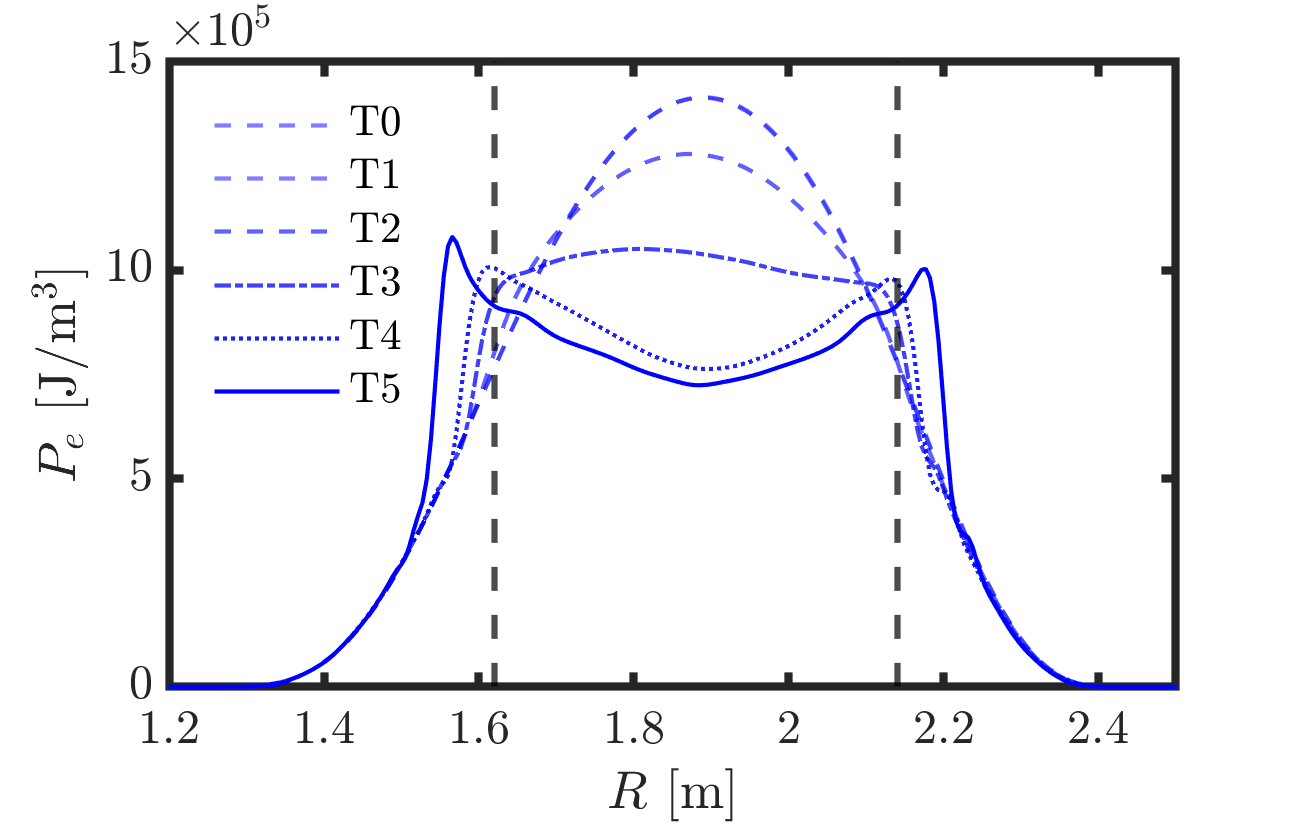}
      (a) $P_e$ profile evolution
      \end{minipage}

      \begin{minipage}[t]{0.48\textwidth}\centering
      \includegraphics[width=\textwidth]{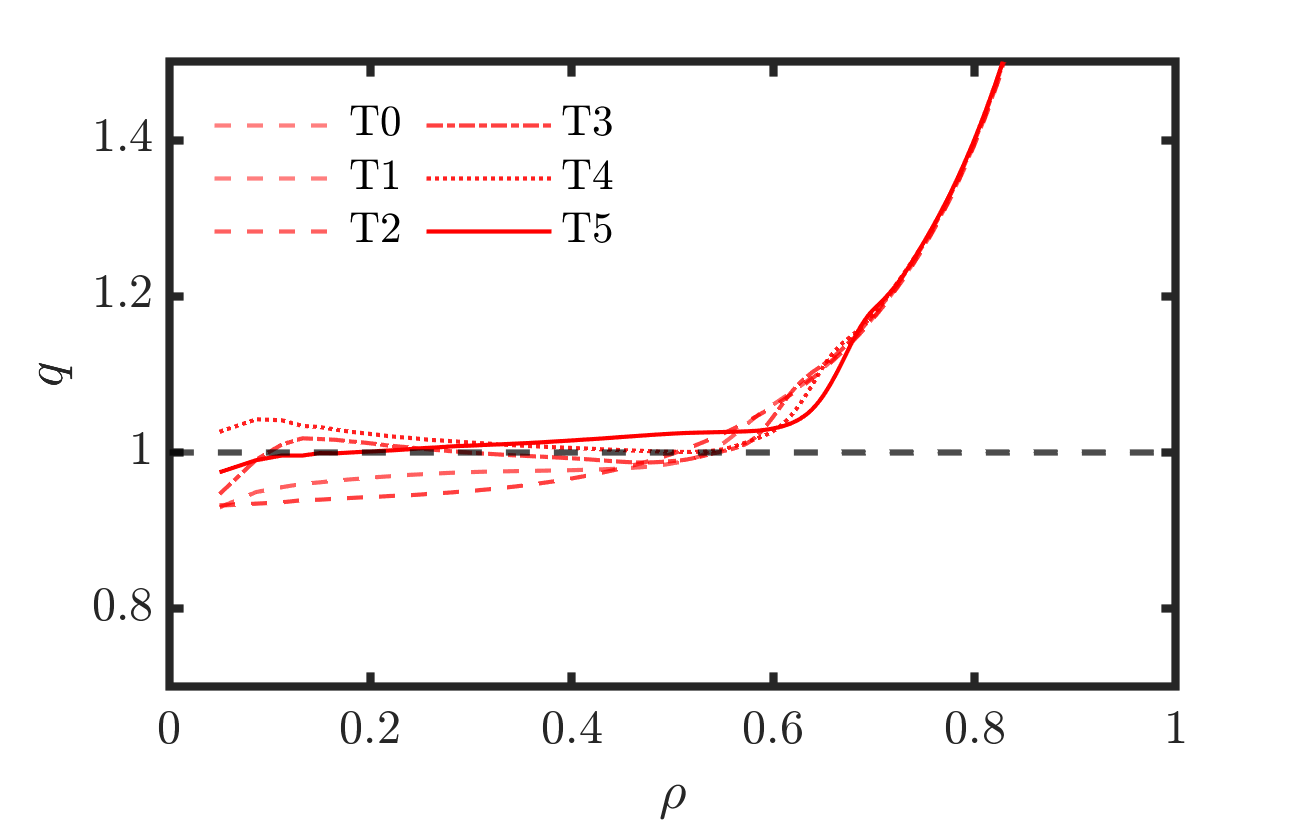}
      (b) $q$ profile evolution
      \end{minipage}

\\

      \begin{minipage}[t]{0.48\textwidth}\centering
      \includegraphics[width=\textwidth]{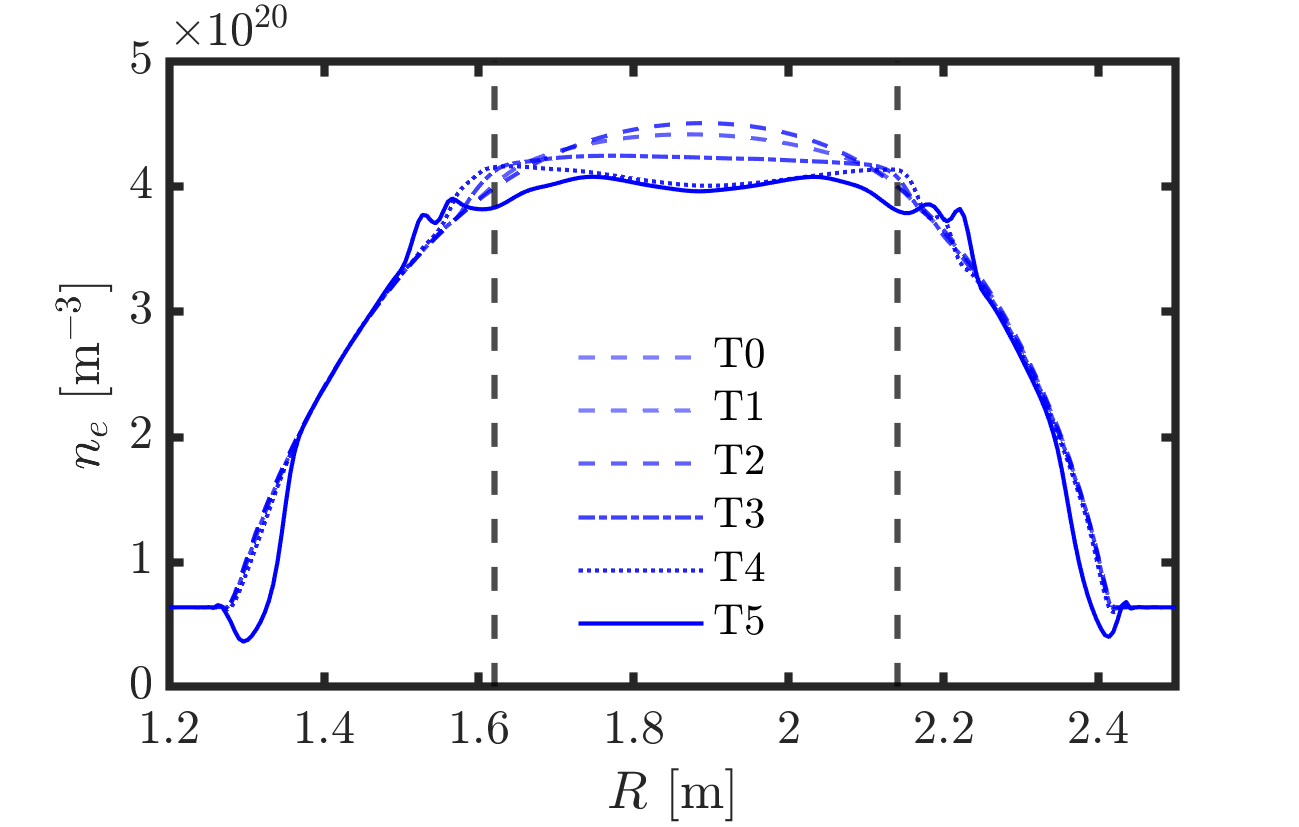}
      (c) $n_e$ profile evolution
      \end{minipage}
      
   \end{tabular}
   \caption{Nonlinear 3D simulation of SPARC baseline case with a thermal conductivity $\bar{\kappa}_T=1.54$ m$^2$/s. The evolution of profiles is consistent with the time slices T1--T5 in Fig.~\ref{fig_nonli_numer_compare}b: (a). electron pressure $P_e$ profile flattening during the first sawtooth crash with the last time slice T5 as a solid line, the initial equilibrium T0 as the faintest dashed line. (b). the safety factor $q$ evolution during the nonlinear stage; (c). the density $n_e$ evolution from T0--T5. The whole device mesh grid is used in this baseline case with toroidal resolution of $n_t=8$ planes.}
   \label{fig_nonli_prd}
\end{figure}
\par
Fig.~\ref{fig_nonli_numer_compare}b shows the time history of kinetic energy spectrum with multiple toroidal mode numbers, $n$ = 0--7 in a nonlinear simulation. In the simulation, we use the whole device mesh (Fig.~\ref{fig_mesh_grid}a) to include the wall effects, so it can give a more self-consistent evolution for SPARC. The required heating and alpha particle physics to drive the equilibrium to this unstable state, as in the previous TRANSP simulations, are missing in our current M3D-C1 simulation, but nevertheless the nonlinear run can provide a rich qualitative analysis for the sawtooth crash flattening process in SPARC.  
\par
Profiles at several representative times are plotted in Fig.~\ref{fig_nonli_prd}a-c to support the following discussion: T1=65~$\mu$s, T2=97~$\mu$s, T3=130~$\mu$s, T4=162~$\mu$s and T5=227~$\mu$s. In the linear stage, a dominant $n=1$ component is observed in the early stage around time T1. As the simulation approaches an initial saturation at T2, higher $n$ modes are excited with larger growth rates roughly 2--4 times the $n=1$ growth rate between T1--T3. The corresponding pressure profile is shown in Fig.~\ref{fig_nonli_prd}a. This early stage results in a less peaked pressure profile.
\par
A very rapid profile flattening occurs during the time range from T3 to T5. Given the equilibrium profiles of the SPARC baseline case, a hollow profile is created. The sharp pressure profile gradient at $R=1.6$~m and $R=2.2$~m introduces a difficulty to numerically solve the fields, and prevents the nonlinear simulation from proceeding for a longer simulation for the baseline case. The numerical tests in Sec.~\ref{subsec:numer_converge} suggests that the sharp pressure gradients can only be resolved with higher-$n$ modes, and that increased toroidal resolution may be required to accurately capture the later nonlinear evolution. In the following discussion, we therefore focus on the initial crash in the baseline simulation.
\par
To better understand the nonlinear evolution, we also plot the corresponding evolution of the $q$ profile (Fig.~\ref{fig_nonli_prd}b) and the density profile (Fig.~\ref{fig_nonli_prd}c). Although a sharp gradient is observed in the pressure profile, the density profile does not have a sharp drop in the present simulation, which implies that the sharp gradient comes from the temperature profile. Some fine structures can also be seen in the density profile at $R=1.3$~m and $R=2.4$~m, which are higher $m$ harmonics around the $q=2$ and $q=3$ locations (similar to the toroidal current perturbation in Fig.~\ref{fig_ntor_scan}b). If these structures exist for a long enough time, other instabilities such as tearing modes may be excited due to the presence of magnetic islands. However, we should keep in mind that the equilibrium we used is a relaxed profile that does not properly describe the edge plasma in SPARC and thus further investigation is needed for edge physics. Moreover, the $q$ profile rises above 1 during the crash, creating a flattened profile around unity near the magnetic axis. This shows that the current profile provides the free energy to drive the $n=1$ instability.
\par
Another key feature is the hollowed profile of the pressure/temperature after the flattening. The destabilizing effects from the high-$\beta_0$ pressure profile can explain this feature as we have shown in the linear simulations. In Fig.~\ref{fig_beta_scan}, the growth rate can increase by $1\sim 2$ orders of magnitude when the baseline case pressure profile is used. The radial mode structure (Fig.~\ref{fig_radial_mode}) shows a large perturbation even on the magnetic axis in the baseline case, as compared to the more localized mode structure in the low-$\beta_0$ case. Based on the 1D model eigenfunction, a circulatory flow is expected to be generated by this strong instability, which can cause the low temperature plasma to flow into the central core region.
\par
To analyze the main cause of the hollowed profile and the mechanism of the initial sawtooth crash in the baseline case, we compare the baseline case with other baseline-like equilibria discussed in previous sections. In these modified cases, we can see how the nonlinear sawtooth crash can be affected by the pressure and the current. An explanation of the observed sawtooth crash is also proposed in the next subsection, inspired by the existing sawtooth crash models.



      


\subsection{Mechanisms for the initial sawtooth crash in SPARC baseline case}\label{subsec:sawtooth_model}


\par
Two major sawtooth crash models have been studied in earlier works: The Kadomtsev model where the central $q_0$ drops below 1 due to the current peaking and a 1/1 resistive reconnection event occurs flattening the plasma profile within $q=1$ surface.\cite{Kadomtsev1975,Sykes1976} The Wesson model described an alternative pressure driven interchange model where a flat $q$ profile very close to unity can lead to an ideal 1/1 interchange instability that flattens the temperature profile on an ideal MHD time scale without a significant change in the magnetic field.\cite{Wesson1986}
\par
\begin{figure}[t]
   \centering
   \begin{tabular}{cc}
      \begin{minipage}[t]{0.48\textwidth}\centering
      \includegraphics[width=\textwidth]{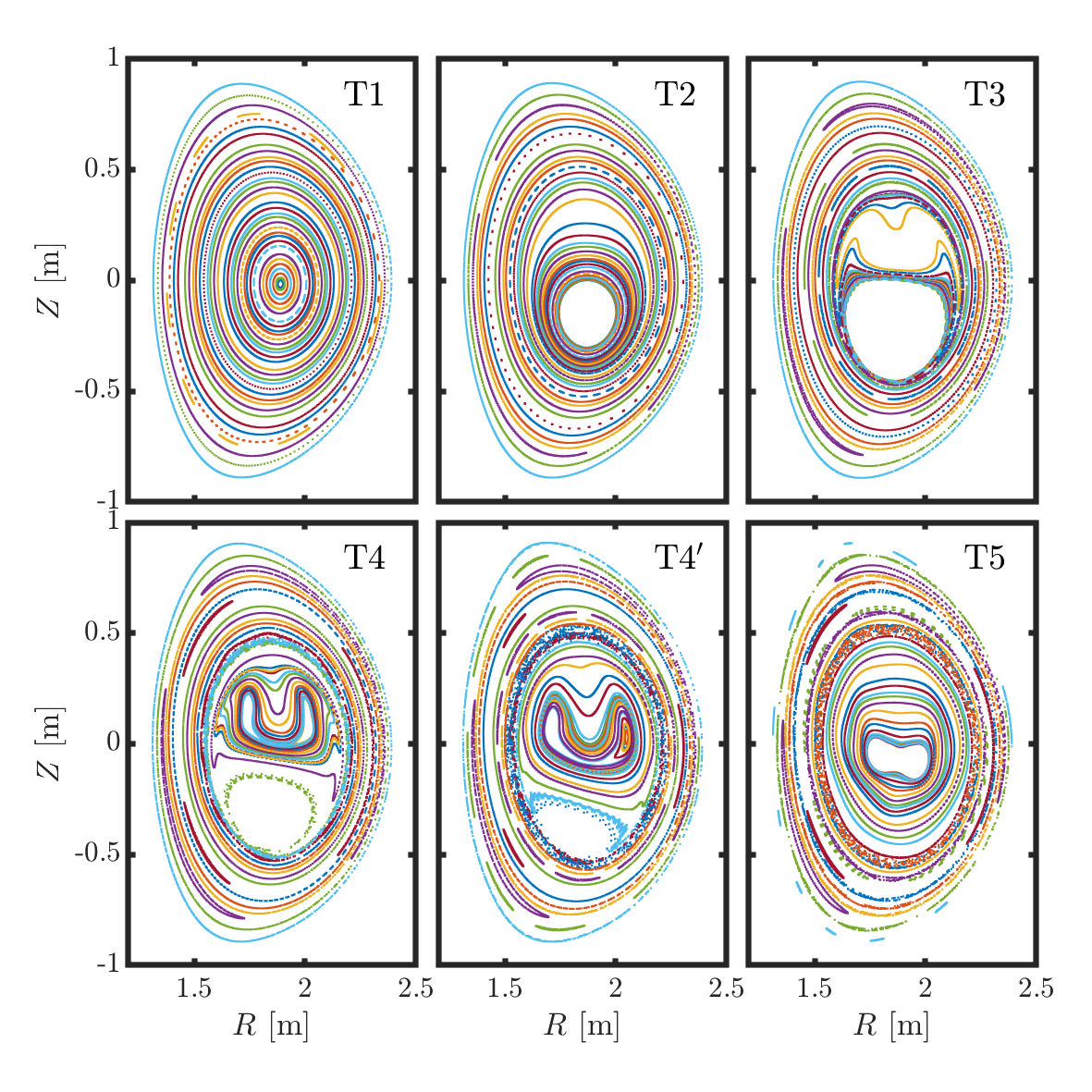}
      (a) Poincar\'{e} plot
      \end{minipage}
      
      \begin{minipage}[t]{0.48\textwidth}\centering
      \includegraphics[width=\textwidth]{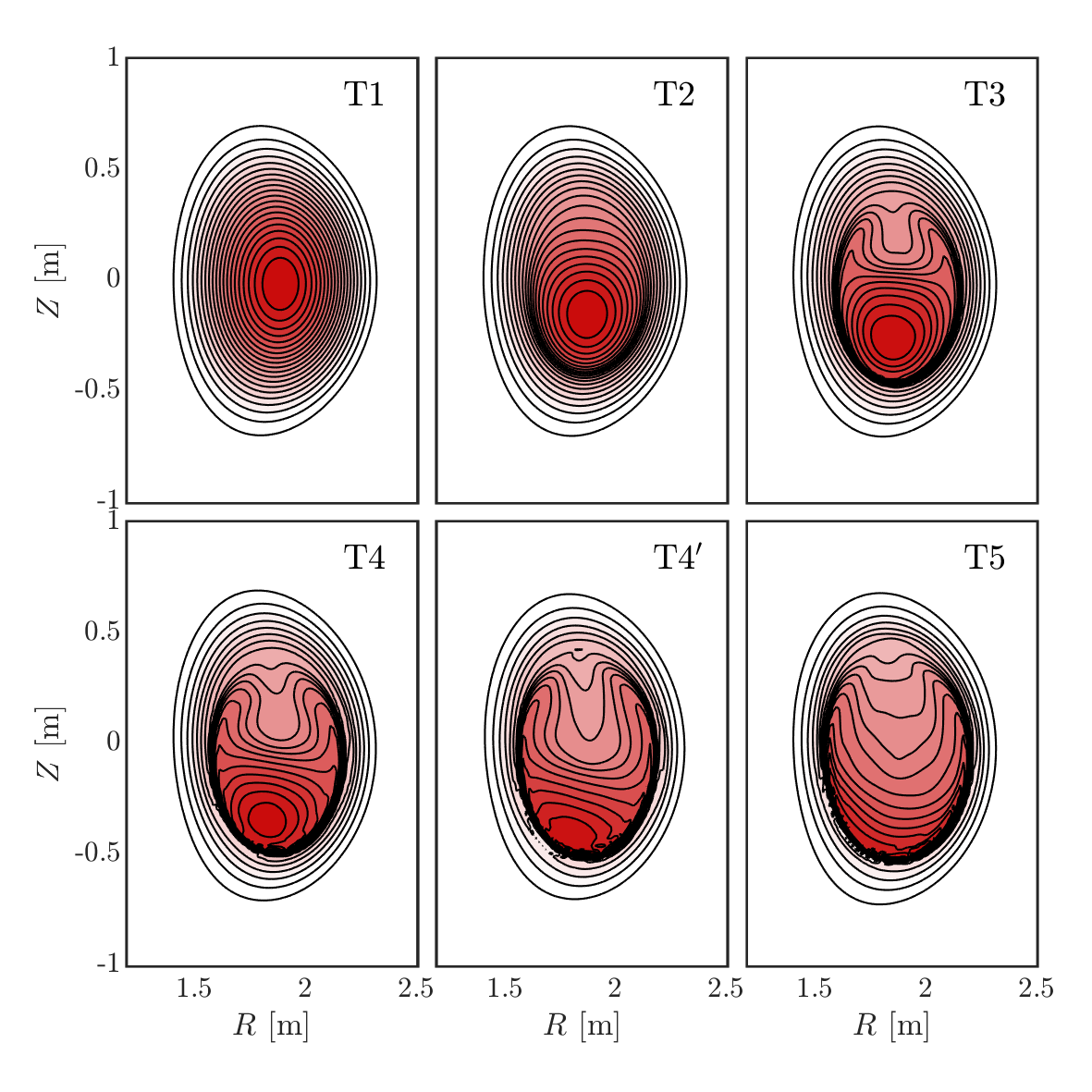}
      (b) Total pressure
      \end{minipage}
   \end{tabular}
   \caption{A sequence of (a) Poincar\'{e} plots of the magnetic field during the initial sawtooth crash for SPARC baseline case, consistent with Fig.~\ref{fig_nonli_prd}. The different colored lines represent different flux surfaces traced out in Poincar\'{e} plots. (b) poloidal plots of the total pressure where the colored contours show the magnitude of the total pressure, all with the same colormap for $P\in(0,2.8\times10^6)$~Pa. The time slices T1--T5 are consistent with Fig.~\ref{fig_nonli_numer_compare}b, and an additional time slice (T4$^\prime$) is inserted in the middle between T4 and T5 time slices, to help illustrate the time evolution of 2D structures.}
   \label{fig_poincare_prd}
\end{figure}

To test if these theories can explain our 3D nonlinear results, we use the Poincar\'{e} plot to show the magnetic field structure in Fig.~\ref{fig_poincare_prd}a for the SPARC baseline case. The time slices correspond to the same ones in Fig.~\ref{fig_nonli_prd}, and an additional time slice, T4$^\prime$=194~$\mu$s, is selected between T4 and T5 to help see the evolution of magnetic field geometry.

\begin{figure}[t]
   \centering
      \begin{minipage}[t]{0.6\textwidth}\centering
      \includegraphics[width=\textwidth]{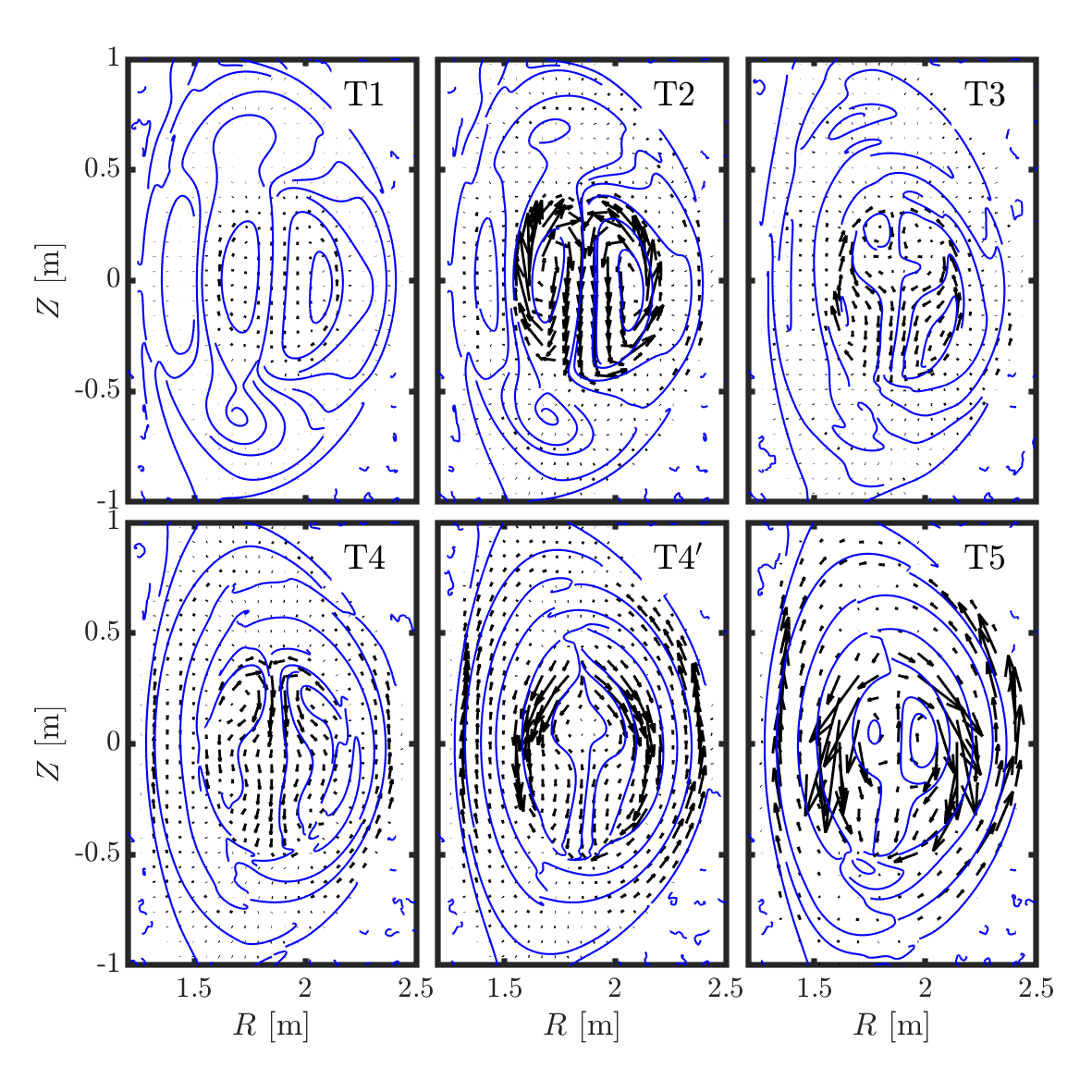}
      \end{minipage}
   \caption{A sequence of poloidal flow fields during the initial sawtooth crash for SPARC baseline case, consistent with Fig.~\ref{fig_poincare_prd}. The contours show representative streamlines. The vectors are scaled based on the actual flow velocities and can be compared across each subplot. The time slices T1--T5 are consistent with Fig.~\ref{fig_nonli_numer_compare}b, and an additional time slice (T4$^\prime$) is inserted in the middle between T4 and T5 time slices.}
   \label{fig_flow_prd}
\end{figure}

\par
The magnetic reconnection event is clear in the Poincar\'{e} plots. At T2, the new magnetic island starts to form near the $q=1$ surface in the upper-half plane and becomes apparent at T3. The old magnetic island gradually shrinks until the new magnetic island with $q=1$ dominates the center of the plasma (T3--T5). The whole process resembles the reconnection process described in Sykes and Wesson (1976)'s 3D nonlinear simulation\cite{Sykes1976}, where the Kadomtsev model is validated in that numerical work.
\par
However, a new feature in our Poincar\'{e} plots compared to these earlier works is that a strongly curved structure begins to grow from T3 to T4$^\prime$. Interestingly, a similar ``island splitting" was observed in Park's 1990 paper\cite{Park1990}. Their TFTR simulation has a very flat $q$ profile around 1 before the crash begins (except very close to the axis due to neoclassical resistivity) and a hollowed profile is observed in the final state.
\par
The flat $q$ profile and the hollowed structure mentioned in Park's work bring our attention to the Wesson model\cite{Wesson1986}. When a flat $q=1$ profile is present, the displacement eigenfunction $\xi_{1,1}$ changes from a rigid shift to a circulatory motion, which can create a hollowed profile resembling a bubble formation process.\cite{Wesson1986} The theory was initially proposed to explain the JET sawtooth data, which holds several inconsistencies with the Kadomtsev model. Based on the X-ray and the electron cyclotron emission (ECE) measurements, a hollowed electron temperature profile is observed during the sawtooth crash.\cite{Edwards1986,Campbell1986,Granetz1988} The role of the poloidal flow induced by the kink mode has also been discussed by Nicolas et al. (2012)\cite{Nicolas2013} for a different observation of a crescent ring structure in the Tore Supra device, where they also try to link the result to the JET data.
\par
Recalling the displacement calculation in the earlier Section~\ref{subsec:beta_model_1d}, a circulatory flow should also be expected in the baseline case. Although the eigenfunction $\xi_{1,1}$ is mainly modified by the high $\beta$ effect instead of a flat $q$ profile, the induced flow should work in the same way as in Wesson's model, and is responsible for the hollowed pressure profile. 
\par
By plotting the time evolution of the total pressure in Fig.~\ref{fig_poincare_prd}b and the flow field in Fig.~\ref{fig_flow_prd}, a clear interchange-type process can be seen. At T2, a clear flow pattern that goes through the magnetic axis can be observed. A ``bubble" starts to grow in the upper-half plane in the T3 pressure plot, consistent with the curved structure observed in the Poincar\'{e} plot and flow field. The ``island splitting" in the following time slices can be associated with this interchange process. The rapid formation of the hollowed structure from T4 to T5 can also be identified in the flow field.


\par
Although there are similarities to the Wesson model, we should keep in mind that the dominant mode in our baseline simulation is a kink mode instead of a direct quasi-interchange mode due to a flat $q\sim 1$ profile. To separate the contribution from the current drive and the pressure drive, we plot the Poincar\'{e} plots and the total pressure contours for constructed cases: (a)~$q_0=0.93$, $\beta_0=0.052$\% and (b)~$q_0=0.991$, $\beta_0=4.75$\%.

\begin{figure}[t]
   \centering
   \begin{tabular}{cc}
      \begin{minipage}[t]{0.48\textwidth}\centering
      \includegraphics[width=\textwidth]{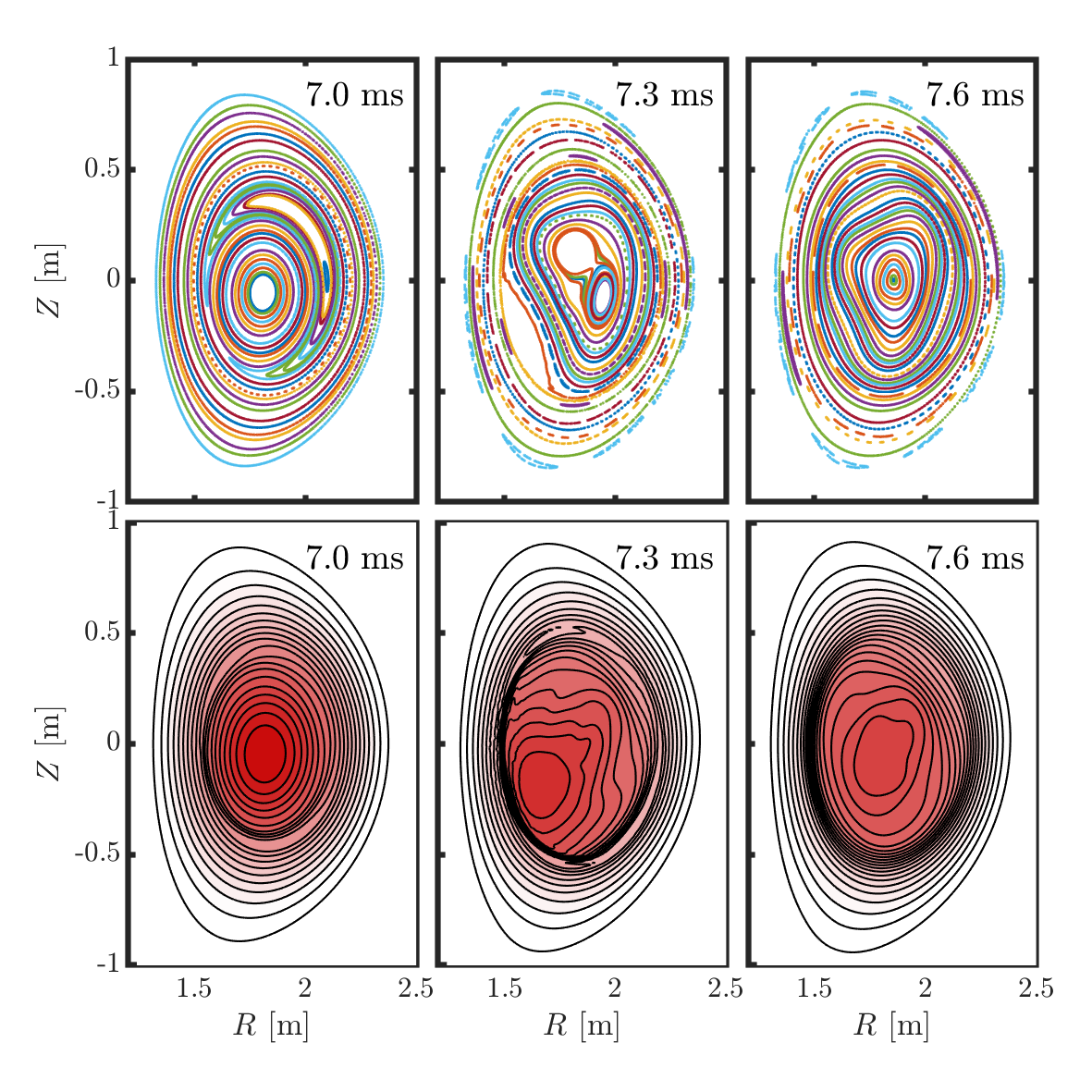}
      (a) $q_0=0.93$, $\beta_0=0.052$\%
      \end{minipage}
      
      \begin{minipage}[t]{0.48\textwidth}\centering
      \includegraphics[width=\textwidth]{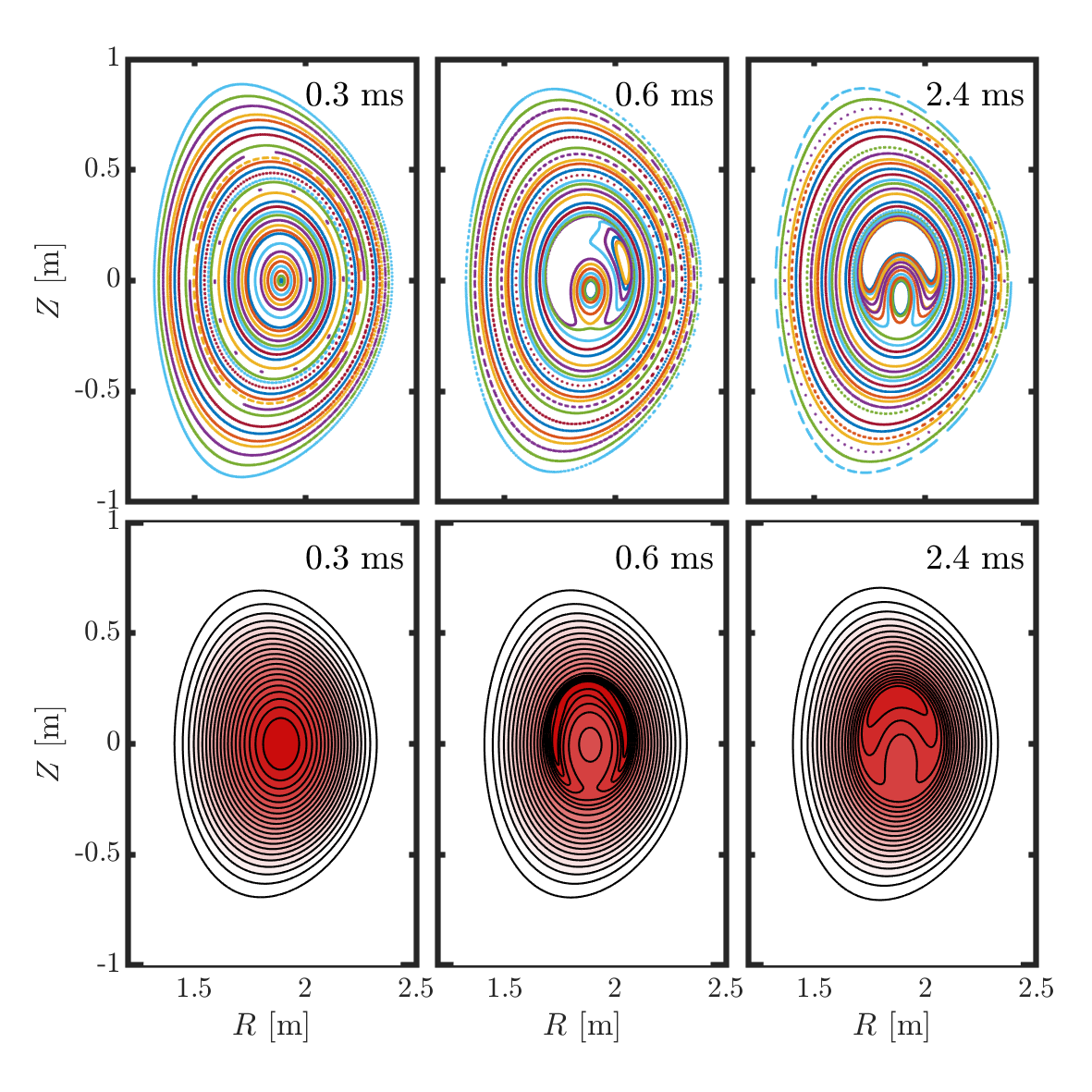}
      (b) $q_0=0.991$, $\beta_0=4.75$\%
      \end{minipage}
   \end{tabular}
   \caption{Sequences of Poincar\'{e} plots of the magnetic field (upper row) and poloidal plots of total pressure (lower row) are compared for the initial sawtooth crash in two cases: (a) with heat conductivity $\bar{\kappa}_T=1.54\times 10^{3}$ m$^2$/s, we obtain the initial condition of $q_0=0.93$, $\beta_0=0.052$\% (Fig.~\ref{fig_kappa_scan}b); (b) with Bateman scale $F_s = 1.07$, we obtain the initial condition of $q_0=0.991$, $\beta_0=4.75$\% (Fig.~\ref{fig_q_shift}a). The different colored lines in a Poincar\'{e} plot represent different flux surfaces traced out in Poincar\'{e} plots and the colored contours in a pressure plot show the magnitude of the total pressure, all with the same colormap for $P\in(0,2.8\times10^6)$~Pa. The time slice is labeled in milliseconds which can be compared to the corresponding cases in previous sections.}
   \label{fig_2d_sequence}
\end{figure}
\par
In Fig.~\ref{fig_2d_sequence}a, we kept the $q$ profile similar to the baseline case but the temperature profile is reduced to $240$~eV to reduce the pressure contribution. As shown in the Poincar\'{e} plots, a typical magnetic reconnection process consistent with the Kadomtsev model can be observed, in which the new magnetic island replaces the old magnetic island. The pressure is flattened without a strongly hollowed structure, although the deformed contour still leads to weakly separated island structures at $t=7.3$~ms, similar to Park's simulation.\cite{Park1990}
\par
In Fig.~\ref{fig_2d_sequence}b, the $q$ profile is lifted to have $q_0=0.991$ so that a relatively flat $q\sim 1$ profile is present within $\rho<0.27$. A 1/1 mode similar to the Wesson model appears, in which an incomplete magnetic reconnection is visualized in Poincar\'{e} plots and a hollowed profile is obtained due to the interchange instability. A postcursor rotation of the mode structure is also seen in the longer simulation after $t=2.4$~ms, though not plotted in this paper. The sensitivity to the $q_0$ value as discussed in Sec.~\ref{subsec:q_scan} (the quick drop of linear growth rate from $q_0=0.991$ case to $q_0=1.008$ case) also agrees with the quasi-interchange model. All these observations well match Wesson's discussion on explaining JET data.

\begin{figure}[t]
   \centering
      \begin{minipage}[t]{0.6\textwidth}\centering
      \includegraphics[width=\textwidth]{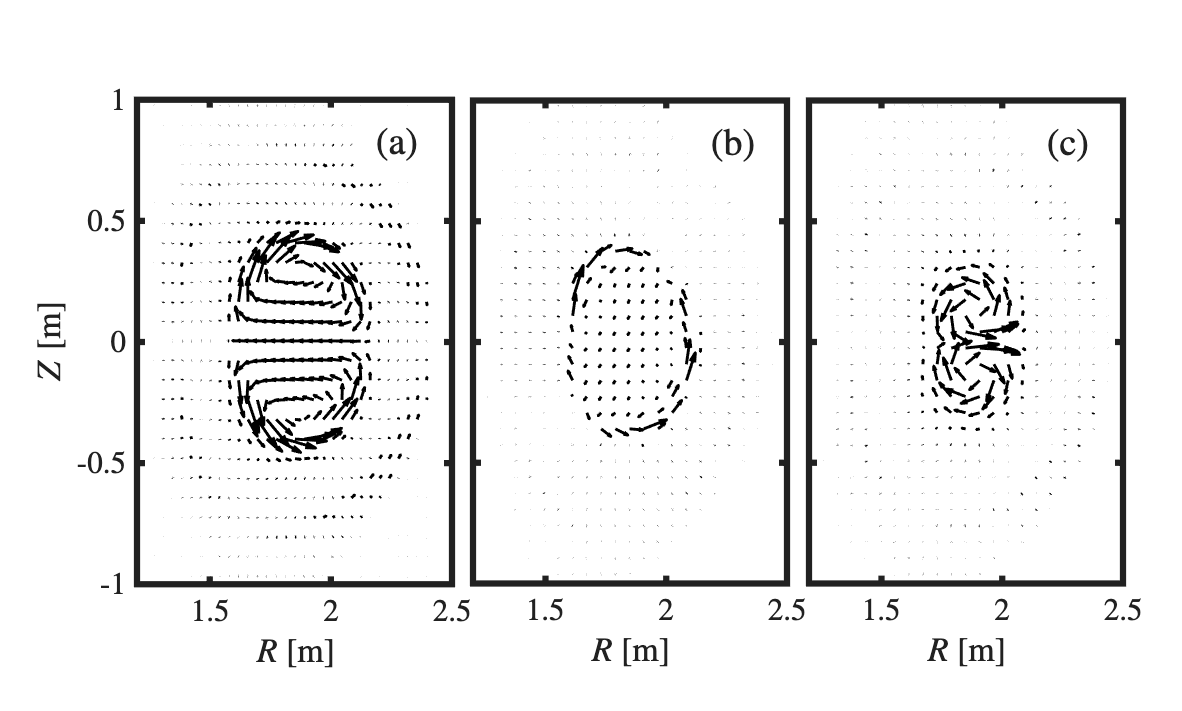}
      \end{minipage}
   \caption{Comparison of poloidal flow field based on linear mode structures for (a). $q_0 = 0.93$, $\beta_0 = 4.75$\%; (b). $q_0 = 0.93$, $\beta_0 = 0.052$\%; (c). $q_0 = 0.991$, $\beta_0 = 4.75$\%. The vectors are auto scaled to show the flow pattern and the relative amplitude within the mode structures.}
   \label{fig_flow_linear_compare}
\end{figure}
\par
At this point, a more complete picture of the initial sawtooth crash in the baseline case emerges, with contributions from both strong current drive and strong pressure drive. To further support this interpretation, the flow patterns of the linear modes are shown in Fig.~\ref{fig_flow_linear_compare}. With current drive dominant, a clear kink mode structure is localized near $q=1$ surface (Fig.~\ref{fig_flow_linear_compare}b). With pressure drive dominant, an interchange structure with strong flow across the magnetic axis is observed (Fig.~\ref{fig_flow_linear_compare}c). In the baseline case, with both current and pressure drives present (Fig.~\ref{fig_flow_linear_compare}a), the mode exhibits a combined feature from both kink and interchange modes, leading to the strong profile flattening and the inflow toward the magnetic axis. A summary of all the nonlinearly simulated scenarios is provided in Table~\ref{tab_nonli_compare}, showing the peak mode growth rate in the linear phase of each nonlinear simulation, the magnetic reconnection type and the pressure drop feature during the sawtooth crash.
\par
In summary, the nonlinear sawtooth crash in the baseline case can be described as below: due to the presence of the $q=1$ surface, an ideal (low-resistivity near $q=1$) n=1 kink mode nonlinearly grows up to perturb the magnetic flux surfaces. Due to the strong $\beta$ effects induced by the high temperature profile, a rotational flow is associated with the kink mode which triggers an interchange-type motion of plasma that can eventually create a hollowed pressure structure. At the end of the sawtooth crash, the new magnetic island with $q_0>1$ overtakes the old magnetic island. Then, plasma reaches the heating phase where Ohmic heating, alpha heating or any other heating mechanisms will continue to change the pressure and current profiles until the second sawtooth crash occurs. This suggests that for high temperature, high current tokamaks like SPARC, the trigger of a sawtooth crash event can be a combined effect from both the current drive and the pressure drive. Assessing the sawtooth oscillation in the PRD case as well as other scenarios of interest is a subject of our future research. More constructed equilibria with different combinations of pressure and $q$ profiles should be evaluated for SPARC, in order to achieve high performance operations. 


\begin{table}[htbp]
  \centering
  \caption{Comparison of nonlinear simulation parameters for sawtooth crashes}
  \label{tab_nonli_compare}
  \begin{tabular}{p{2.5cm} p{2cm} p{2cm} p{2cm} p{2cm} p{2cm}}
    \toprule
    \textbf{($q_0,\beta_0)$}
    & \textbf{$(0.93,4.75\%)$} 
    & \textbf{$(0.991,4.75\%)$} 
    & \textbf{$(1.008,4.75\%)$} 
    & \textbf{$(0.93,0.052\%)$} 
    & \textbf{$(0.93,0.009\%)$} \\
    \midrule
    \textbf{Mesh geometry} & Whole device & Simple  & Simple  & Simple  & Simple  \\
    \addlinespace
    \textbf{Toroidal $n_t$} & 8 & 4 & 4 & 4 & 4 \\
    \addlinespace
    \textbf{Linear phase $\gamma\tau_A$} & $1.67\times 10^{-2}$ & 
    $1.07\times 10^{-2}$ & 
    $8.09\times 10^{-4}$ & 
    $9.38\times 10^{-4}$ & 
    $5.01\times 10^{-3}$ \\
    \addlinespace
    \textbf{Reconnection type} & Full reconnection & Partial reconnection & Negligible, distorted flux & Full reconnection & Full reconnection \\
    \addlinespace
    \textbf{Pressure drop} & 49\% of $P_0$, hollowed profile & 22\% of $P_0$, hollowed profile & 5\% of $P_0$ & 27\% of $P_0$ & 20\% of $P_0$ \\
    \bottomrule
  \end{tabular}
\end{table}

\subsection{Discussion on sawtooth oscillation timescales}\label{subsec:sawtooth_timescale}

\par
In the low-temperature simulations with artificially enhanced $\kappa_T$, the electron pressure profile shows an oscillating behavior, which resembles sawtooth oscillations. In this section, we analyze several characteristic timescales as an initial effort to understand sawtooth oscillations in SPARC. However, because we use the simple geometry mesh with $n_t=4$ in these simulations, the timescale estimations should be treated only as order-of-magnitude values. More importantly, the simulated temperatures are far below the SPARC operating parameters, and several important physical effects, such as two-fluid effects, kinetic alpha particles and associated heating sources, are missing in our current simulations. The timescales discussed here should, therefore, not be interpreted as quantitative predictions for SPARC operating scenarios. 
\par
As labeled from T1 to T3 in Fig.~\ref{fig_kappa_scan}a, the pressure profile will periodically peak and flatten during a cycle, with a period of 0.2~ms. To estimate the crash time, we take the inverse of the estimated growth rate in the nonlinear simulation, that during the first crash, $1/\gamma = 0.036$~ms, and the subsequent crash, $1/\gamma = 0.032$~ms. The $q$ profile evolution (third plot) shows a large change during the first sawtooth crash in the last panel of the figure. The $q_0$ value rises above 1 and later stays close to 1 during the following profile oscillations. 
\par
The simulation with $\bar{\kappa}_T=1.54\times 10^{3}$ m$^2$/s shows a similar sawtooth oscillation. With a lower heat conductivity, the input profiles reached a new balance with a peaked electron temperature $T_{e,0} = 240$~eV ($\beta_0 = 0.052$\%). Similar sawtooth oscillations are observed after the initial crash, except now with a much longer period of 1--1.5 ms, as shown in Fig.~\ref{fig_kappa_scan}b. To estimate the crash time, we take the inverse of the growth rate, that during the first crash, $1/\gamma = 0.22$~ms, and the subsequent crash, $1/\gamma = 0.081$~ms. The $q$ profile is again lifted up above 1 in the first crash, and then oscillates around unity in the following MHD bursts. 
\par
A clear timescale separation is observed between the sawtooth oscillation period and the mode growth (about one order of magnitude). For the high $\beta_0=4.75$\% in the baseline case, we may expect a much longer period and relatively fast crash time for the second sawtooth oscillation. This may explain why a similar oscillation is not observed yet in the pressure-driven sawtooth crash (Fig.~\ref{fig_q_shift}) within our simulation time. 
\par
The pressure profile, however, does not restore to its original on-axis value before the first sawtooth crash. After the initial profile flattening, the pressure oscillates in a relatively narrow range around the flattened profile (about 70\% of its magnitude before crash). The electron pressure is limited to the flattened profile in the following sawtooth oscillations. A clear increase of the growth rate (roughly 3 times) after the initial crash may also suggest a different process for the following sawtooth oscillation in the $\beta_0=0.052$\% case. Further analysis is needed to fully understand why the pressure profile fails to peak again after the initial sawtooth crash.
\par
The discussion in this paper focuses on the initial sawtooth crash in our simulations, since the periodic bursts after the initial crash behave differently. Especially, the $q$ remains in a relatively flat profile close to unity in the later crashes. This resembles the pressure-driven sawtooth crash with a sensitive dependence on $q$ near unity described by Jardin, et al.\cite{Jardin2020}, but no rigorous comparison has been made yet. Other physics can also be important for a proper estimation of the restoration timescale, including heating mechanisms such as alpha heating and ICRH heating. The two-fluid effects or kinetic effects can also play a role in stabilizing or destabilizing the mode, which should be evaluated in the future work.
\par
Another limitation of our simulations is that the Ohmic heating is the main heating source in our simulations. A sufficiently long sawtooth oscillation period is expected to restore the peaked profile by the Ohmic heating only. In the designed $Q=11$ PRD scenario, the ICRH heating power is $P_{RF}=11.1$~MW, the Ohmic heating power is $P_{Ohmic}=1.7$~MW and the estimated alpha heating power is $P_\alpha\sim 28$~MW out of $P_{fusion}=140$~MW.\cite{Creely2020} Most of these heating sources, which are important to reproduce the self-consistent sawtooth oscillation, are absent in our current simulations. In addition, TRANSP simulations suggest a characteristic timescale for the sawtooth oscillation period of around $1$~s \cite{Rodriguez-Fernandez2020}, which can be computationally challenging for  the nonlinear 3D simulations. Therefore, a quantitative discussion of the sawtooth oscillation period is beyond the scope of this work and is subject to future investigations.

\par
\quad 

\section{Conclusion}\label{sec:conclusion}


In this paper, we carried out a comprehensive analysis for low-$n$ MHD instabilities in several SPARC baseline-like equilibria. Limited by the single-fluid extended-MHD model with Ohmic heating in our present simulations, we emphasize the analysis of the potential sawtooth crash mechanisms that may appear in SPARC. This will be a basis for the future work on quantitative predictions of sawtooth oscillations in SPARC.
\par
The equilibrium used in this paper is a relaxed profile without a pedestal region, featuring a peaked temperature of $20$~keV, comparable to the PRD scenario for SPARC. The on-axis $q_0$ is $0.93$, which allows an unstable $m=1$, $n=1$ internal mode to grow in the core region. 
\par
Linear simulations suggest that this 1/1 mode is the most unstable mode in the baseline case. The mode can be identified as an internal kink mode by scanning over the Lundquist number $S$. Both the tearing-like scaling ($\gamma \propto S^{-0.56}$, within $S\in [3.35\times 10^4,1\times 10^{6}]$) and the resistive-kink-like scaling ($\gamma \propto S^{-0.35}$, within $S\in [1\times 10^4,2.51\times 10^{4}]$) can be verified in the \simcode{} simulations and qualitatively agree with the theoretical discussion by Hastie (1987)\cite{Hastie1987}. The SPARC PRD design parameter near $q=1$ surface gives a large $S>10^9$, which sets the observed kink mode close to its ideal MHD limit.
\par
Both the current drive and the pressure drive play important roles in explaining the strong kink mode in the SPARC baseline case. The 1/1 mode growth rate can be greatly reduced by an order of magnitude when the on-axis plasma $\beta_0$ drops below $2$\% ($\sim$~keV), or when the $q_0$ approaches unity. The combined effects of strong current drive with low magnetic shear and a steep pressure gradient significantly enhance the mode growth rate.
A broader radial mode structure is associated with the high-$\beta_0$ profile, and it becomes more localized around $q=1$ when lowering the $\beta_0$. These observations can be qualitatively reproduced in a 1D screw pinch model for a single $m$, single $n$, by matching the density perturbation between the 1D model and the \simcode{} result.
\par
We begin our nonlinear simulation from the marginally unstable cases. We reduce the current drive by shifting the $q$ profile towards 1 while keeping the pressure drive unchanged (fixed $\beta_0 = 4.75$\%). As $q_0$ changes in a narrow range around unity (from $0.991$ to $1.008$), the $n=1$ mode is quickly reduced, suggesting a sensitive behavior of $n=1$ mode with $q_0\sim 1$. In the simulation with $q_0=0.991$, the hollowed structure can still be observed, which implies that the hollowed structure in the baseline simulation is mainly induced by the pressure drive (high temperature). 
\par
By lowering the temperature while keeping $q_0$ around 0.93, we can remove most of the pressure drive while maintaining the current drive. Sawtooth oscillations are observed in nonlinear runs with low $\beta_0$ values. 
For the two cases we tested ($\beta_0=0.009$\% and $0.052$\%), the pressure is weakly oscillating without a hollowed structure after the initial crash, and the $q$ profile stays close to 1 during the later stage multiple bursts. The limited pressure profile oscillation after the initial crash and the sawtooth oscillation period in our simulations need further investigations. This can be a more sophisticated task that involves the plasma profiles, the heating sources, and the triggering mechanisms, which are part of the ongoing tasks in our future work. 
\par
Moving back to the baseline case, both the pressure and the current are strongly destabilizing the plasma. The PRD design point reaching such a low-$q_0$, high-$\beta_0$ equilibrium likely requires multiple stabilizing effects to postpone the onset of the 1/1 instability in SPARC. Although auxiliary heating and kinetic alpha particles are not included, our nonlinear simulations nevertheless provide a rich qualitative understanding of the self-consistent profile flattening in SPARC. After a linear growth phase with a dominant $n=1$ component, the mode nonlinearly saturates as higher toroidal components grow up, eventually leading to a drastic profile flattening with hollowed pressure and temperature profiles. The strong crash is a combined result of both the current profile and the pressure profile in the baseline case.
\par
Inspired by the Kadomtsev model and the Wesson model, the baseline case can be explained as an internal kink triggered sawtooth crash, strengthened by a pressure-driven interchange-type motion. The process includes a clear magnetic reconnection event, and at the same time an associated circulatory flow creates a hollowed pressure profile. To further support this interpretation, both the Kadomtsev model and the Wesson model can be verified individually by looking at the modified simulations that separate the current and pressure drives, as we discussed above (modifying $\beta_0$ and $q_0$ based on the baseline case).
\par
This study is a first step for understanding the confinement of particles during sawtooth oscillations in SPARC. The particle confinement can be crucial since good confinement of high energy particles, such as alpha particles, is necessary for heating in high-$Q$ operation of SPARC.
\par
More efforts are needed to capture a complete picture of nonlinear evolutions of sawtooth oscillations in SPARC. The inclusion of multiple heating sources and alpha particles will be subject to future research. In current \simcode{} simulations, the Ohmic heating is the only heating source, which can take a few hundred milliseconds or even seconds to restore the pressure profile after the initial sawtooth crash. Additional heating sources, such as ICRH heating, alpha particle heating and additional loop voltages, should be considered to check if more frequent keV-level sawtooth oscillations can happen in SPARC. 
\par
Considering the strong pressure effects on the kink mode in SPARC, the two-fluid effects and other kinetic effects can strongly modify the stability properties in SPARC. The kinetic alpha particles can also stabilize the kink mode in these high-temperature tokamaks. Modeling this effect requires a hybrid kinetic-MHD model for simulations. It can be an interesting task to explore if we can predict the sawtooth oscillation period and the magnitude of crashes, or test theories to trigger or prevent sawtooth events in SPARC. However, we should keep in mind that quantitatively predicting sawtooth activity, including both the trigger mechanism and the resulting crash dynamics, remains a major challenge. This comes from its inherent nature of large temporal and spatial scale separation and the coupling of multiple physical processes, including auxiliary heating, two-fluid and kinetic effects, and energetic alpha particle dynamics.
\par
More SPARC scenarios should be explored using the nonlinear \simcode{} simulations to validate the present results. Exploration of equilibria with different combinations of pressure and $q$ profiles will be considered in future work. A near future goal of SPARC, to achieve a $Q>1$ scenario with the help of the ICRH heating, can be a good test case to continue the sawtooth oscillation simulation presented in this paper. Including a realistic heating source in our simulations can also provide insights on predicting potential sawtooth oscillations in the SPARC tokamak.


\ack{

The authors would like to thank Dr. Dylan P. Brennan (General Fusion Corp.), Yiru Xiao (MIT), and  Dr. A. Kumar (MIT) for  solving the 1D model to analyze plasma $\beta$ effects in linear simulations for kink modes. Dr. R. Granetz (MIT), Dr. R. Chandra (MIT) and Dr. A. Kumar (MIT) provided great insights for completing the discussion on nonlinear simulations of sawtooth crashes. This work is supported by the Commonwealth Fusion Systems (CFS), grants \#RPP020 and \#RPP023. The simulations use resources on the National Energy Research Scientific Computing Center (NERSC), grant \#10.13039/100017223, and the MIT Engaging cluster at the MGHPCC facility (https://engaging-web.mit.edu).

}


\data{
 \simcode{} is now an open-source code: https://github.com/PrincetonUniversity/M3DC1. All simulation input files and published result data in this paper are available upon reasonable request.
}


\bibliographystyle{unsrt} 
\bibliography{references}  

\end{document}